\documentclass[11pt]{article}
\usepackage{graphicx}
\usepackage{amsmath, amssymb}
\usepackage{natbib}
\usepackage{hyperref}
\usepackage{booktabs,tabularx, colortbl}
\usepackage{authblk} 
\usepackage{color} 
\usepackage{tcolorbox}
\usepackage{enumitem}
\usepackage{arydshln}
\usepackage{cleveref}

\usepackage[left=2.5cm,right=2.5cm,top=3cm,bottom=3cm,centering]{geometry}

\renewcommand\footnotemark{}

\title{\centering Deselection of Base-Learners for Statistical Boosting - with an Application to Distributional Regression}

\author[1]{Annika Strömer\thanks{This article is not an exact copy of the original published article in \textit{Statistical Methods in Medical Research}. The definitive publisher-authenticated version is available online at: \url{https://doi.org/10.1177/09622802211051088}.\\ \hspace*{4mm}
\textit{Corresponding author:} Annika Str\"omer, Department of Medical Biometrics, Informatics and Epidemiology,  Faculty of Medicine, University
of Bonn, Venusberg-Campus 1, 53127 Bonn, Germany,}}
\author[1]{Christian Staerk}
\author[3]{Nadja Klein}
\author[1]{Leonie Weinhold}
\author[2]{Stephanie Titze}
\author[1]{Andreas Mayr}

\small{
\affil[1]{ Department of Medical Biometrics, Informatics and Epidemiology, University
of Bonn, Germany}
\affil[2]{ Department of Nephrology and Hypertension, FAU Erlangen-Nuremberg, Germany}
\affil[3]{ Humboldt-Universit\"at zu Berlin, Berlin, Germany}
}

\date{} 

\begin{document}
\setlength{\parindent}{0em} 
 
\maketitle

\vspace*{-10mm}

\begin{abstract}
We present a new procedure for enhanced variable selection for component-wise gradient boosting. Statistical boosting is a computational approach that emerged from machine learning, which allows to fit regression models in the presence of high-dimensional data.  
 Furthermore, the algorithm can lead to data-driven variable selection. In practice, however, the final models typically tend to include too many variables in some situations. This occurs particularly for low-dimensional data ($p<n$), where we observe a slow overfitting behavior of boosting. As a result, more variables get included into the final model without altering the prediction accuracy. Many of these false positives are incorporated with a small coefficient and therefore have a small impact, but lead to a larger model. We try to overcome this issue by giving the algorithm the chance to deselect base-learners with minor importance. We analyze the impact of the new approach on variable selection and prediction performance in comparison to alternative methods including boosting with earlier stopping as well as twin boosting. We illustrate our approach with data of an ongoing cohort study for chronic kidney disease patients, where the most influential predictors for the health-related quality of life measure are selected in a distributional regression approach based on beta regression. 
\end{abstract}

\section{Introduction}
\label{Intro}
In modern biostatistics, model building and variable selection have become increasingly important, particularly in the context of applications in high-dimensional data settings, where the number of potential predictors $p$ is larger compared to the sample size ($p\gg n$)\citep{HighDimData}. Important examples include genetic or molecular data  (e.g., \citealt{jasa_prs};\citealt{choi2020tutorial}), but also in more classical clinical studies one often aims to obtain a relatively sparse model with good prediction accuracy including only the most relevant variables (e.g., \citealt{steyerberg2014towards}; \citealt{sauerbrei2020state}). 

Component-wise gradient boosting~\citep{buhlmann} provides a framework to handle this, with the key features of variable selection and the possibility to manage high-dimensional data problems. In combination with regression-type base-learners~\citep{mayr2017update}, it is able to maintain the usual interpretability of statistical regression models -- equivalent to the ones that were estimated using classical penalized likelihood or Bayesian inference. Statistical boosting provides a large flexibility due to the modular nature of the approach: any type of base-learner (linear models, splines, spatial models) can be combined with any type of convex loss function~\citep{buhlmann2014discussion}. 

Despite these advantages, in some applications the algorithm tends to select too many variables. This often occurs for rather low-dimensional settings with relatively large sample sizes ($p<n$), where irrelevant base-learners often get selected with a very small effect size. This is associated with slow overfitting and thus with a higher number of boosting iterations $m_{\rm{stop}}$ which lead to a larger set of selected variables. For example, in a recent beta-regression analysis on the health-related Quality of Life (QoL) in $n = 3522$ chronic kidney disease patients, statistical boosting selected 33 out of $p = 54$ potential variables~\citep{betaboost}.  

\begin{figure}[t]
\centering
\includegraphics[scale = 0.4]{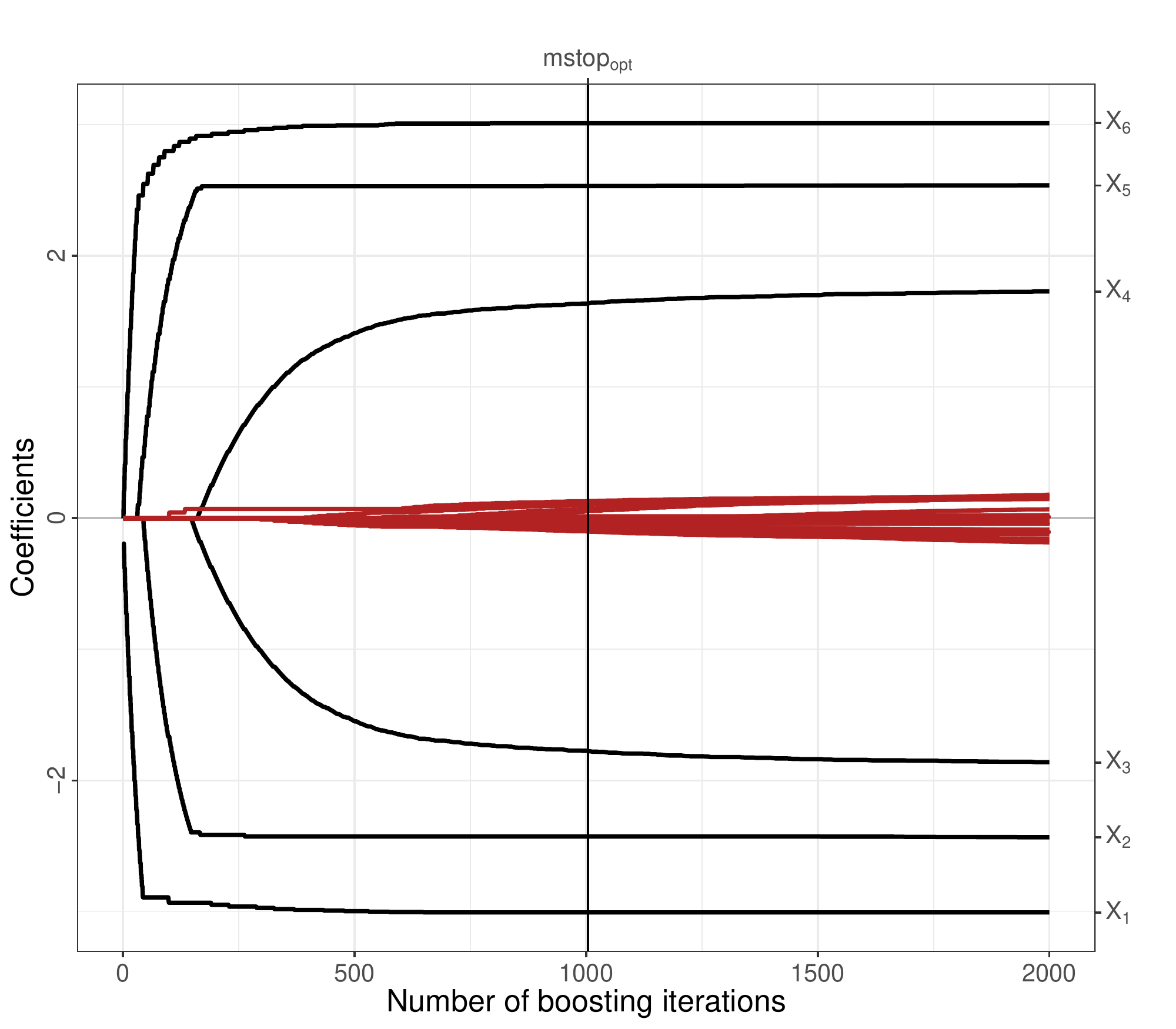}
\caption{Coefficient paths along the number of boosting iterations for a simulated data set with $n = 500$ observations and $p = 50$ variables which were simulated from a multivariate normal distribution with Toeplitz covariance structure and a correlation of 0.9. Only variables $X_1,\dots, X_6$ are \emph{informative} with true coefficients  $\beta_{\text{inf}} = (-3,-2.5,-2,2, 2.5,3)^T$. The coefficient paths for the \emph{non-informative} variables are colored red. The optimal stopping iteration $\text{mstop}_{\text{opt}} = 1004$ was determined by 10-fold cross-validation.  
\label{Figure:Path_Risk}}
\end{figure}

As an illustration, Figure~\ref{Figure:Path_Risk} displays the coefficient paths of component-wise boosting with the squared error loss in the context of linear regression for a simulated data set in which only the first six variables $X_1,\dots, X_6$ are informative. 
One can observe that the estimated coefficients of the six informative variables are the largest in absolute values, while several non-informative variables are incorporated with small coefficient sizes around zero. Therefore, the selected non-relevant variables have only a minor impact on the predictive performance but lead to a larger model with difficult interpretation.  

\cite{twin} tried to overcome this issue and extended the classical boosting approach to a two-stage design called twin boosting, which was inspired by the adaptive lasso (see~\citealt{adaLasso}).
The first stage consists of a classical boosting algorithm. The second stage is similar to the first, with the difference that variables that have not been selected are excluded; furthermore, variables that have been selected in the first stage receive weights based on the size of their coefficients, making it more likely that the important variables will be selected again in the second stage. 
Other approaches aiming to increase the sparsity of statistical boosting focus on  reducing the number of iterations $m_{\rm{stop}}$: for example, the one standard-error rule was originally considered by \cite{breiman} in the context of random forests and does not select the optimal tuning parameter regarding prediction accuracy, but in case of boosting the smallest $m_{\rm{stop}}$ that is still in the margin of one standard error from the minimum risk. \cite{RobustC} further extended this approach (\emph{RobustC}) to obtain a less complex prediction rule that is less affected by the characteristics of the resampling scheme compared to the one standard-error rule. A potential disadvantage of approaches that lead to earlier stopping is that they suffer from the side-effect of inducing also a higher amount of shrinkage. This additional shrinkage of selected effect estimates might not necessarily lead to a better performance (cf.,\citealt{van2020regression}). 

Here we propose a general procedure to enhance the sparsity of statistical boosting models, where the final selection of variables is based on the risk reduction resulting from the individual updates of the corresponding base-learners. 
With this approach, we exclude those base-learners (and their corresponding variables) from the prediction model which attributed only slightly to the total risk reduction. 
As an alternative to earlier stopping of the boosting algorithm -- i.e.\ moving ``horizontally'' on the regularization paths -- we consider the individual contributions of different variables after a particular number of boosting iterations. The benefits of this ``vertical'' view on regularization paths have also recently been discussed in the context of other regularization methods such as the thresholded Lasso (\citealt{zhou2009thresholding, weinstein2020power}) including the possibility of deselecting noise variables which are included ``early'' on the regularization paths. Furthermore, we directly enforce the sparsity of the final models without unnecessarily increasing the amount of shrinkage on effect estimates. 
We illustrate the proposed method with the selection of predictors for the health-related QoL data of the German Chronic Kidney Disease Study (GCKD). We compare our results to a previous analysis of these data~\citep{betaboost} which partly motivated the new methodological development. With the new deselection approach, we are able to select much sparser models while still yielding a similar prediction performance. 

The remainder of the paper is structured as follows. In Section~\ref{VS} we introduce the new approach for an improved variable selection and consider alternative methods for achieving sparser models. In Section~\ref{Simulation} we compare the methods considered in Section~\ref{VS} via simulated data under various conditions for different models. Finally, we apply our new approach to the quality of life data and present the results in Section~\ref{GCKG}. Conclusively, Section~\ref{Discussion} summarizes our findings and discusses future research questions.

\section{Methods}\label{VS}
\subsection{Model-based boosting}\label{ClassicalBoosting}

Boosting was first established in the context of machine learning (\citealt{Boosting_ML, FreundShapire}) and was later extended to fit statistical models (\citealt{friedman2000,friedman2001}). Statistical boosting algorithms (\citealt{mayr2014evolution, mayr2014extending}) can be used to analyze high-dimensional data problems, in which classical inferential methods are no longer applicable (e.g. least squares method for linear regression models). Furthermore, boosting yields data-driven variable selection and shrinkage of effect estimates~\citep{buhlmann}.

The model fitting is carried out by iteratively minimizing the empirical risk of an appropriate loss function. This loss defines the regression problem and needs to be specified in advance.  In generalized linear models (GLMs) and generalized additive models (GAMs), the loss function corresponds to the negative log-likelihood of the outcome distribution.  For classical linear regression models, for example, we minimize the squared error ($L_2$ loss), which corresponds to maximizing the likelihood of a Gaussian distribution.  
Different effect types can be determined for each covariate (e.g. linear or smooth effects), which reflect the type of influence the variable has in the model. These underlying functions are called base-learners; in the simplest case they are univariate linear models representing linear effects. 
In each iteration, the negative gradient of the loss function is determined and every base-learner is separately fitted to the negative gradient. Afterwards, only the best performing base-learner is selected (i.e. the base-learner that best fits the negative gradient) and the corresponding estimated effect is multiplied by a small fixed step size (default is $\nu = 0.1$) before it is included in the model.
Due to the selection of single base-learners in each iteration, the algorithm carries out variable selection. This process is repeated until the number of boosting iterations $m_{\rm{stop}}$ is reached, whereby every base-learner can be selected several times. In the classical boosting algorithm, every base-learner that was once included in the model can not be deselected~\citep{RTutorial}.

The number of boosting iterations is the main tuning parameter and can be selected e.g. by cross-validation or other resampling techniques. The optimization of the stopping iteration -- also referred to as \textit{early stopping} --  is crucial to prevent overfitting and to favor the sparsity of the resulting model. The smaller $ m_{\rm{stop}}$, the fewer variables are included in the final model as only one base-learner is updated in each iteration. Additionally, early stopping typically improves the prediction accuracy and leads to shrinkage of effect estimates~\citep{knowstop}.

\subsection{Earlier stopping strategies}\label{ES}
Due to the influence of the number of boosting iterations  $ m_{\rm{stop}}$ on the variables finally selected by the algorithm, one approach to achieve sparser models is to enforce \textit{earlier} stopping of the algorithm, i.e. selecting a smaller  $m_{\rm{stop}}$. With this approach it is assumed that variables that are updated in early iterations of the algorithm have a greater influence on the prediction of the model than variables that are added later to the model. Typically, classical early stopping selects the stopping iteration $m_{\text{stop\_opt}}$ that leads to the smallest (optimal) cross-validated prediction risk (CV).

The one standard error rule (oSE) is one approach to enforce earlier stopping and has already been used in context of penalized regression and regression trees~\citep{breiman,glmnet}.
With this approach, the tuning parameter $m_{\rm{stop}}$ is chosen as the smallest iteration for which the CV is within one standard error of the minimal CV (cf. \citealt{glmnet}; \citealt{hastie2009elements}):
\begin{equation*}
\text{CV}(m_{\rm{stop}}) \leq \text{CV}(m_{\rm{stop \_ opt}}) + \text{se}(\text{CV}(m_{\rm{stop \_ opt}})).
\end{equation*}
The minimal cross-validated predictive risk $\text{CV}(m_{\rm{stop \_ opt}})$ corresponds to the CV of the optimal stopping iteration. Furthermore, $\text{se}(\text{CV}(m_{\rm{stop \_ opt}}))$ represents the standard error of the minimum over the CV folds. Consequently, this method has dependencies on the number of CV folds and the sample size.

Based on the idea of the oSE approach, \cite{RobustC} proposed an alternative more robust approach, called RobustC. 
Here the smallest $m_{\rm{stop}}$ is chosen, whose CV is still within a range of a fixed additional tuning parameter $c_{rC}$ multiplied with the minimum CV:
\begin{equation*}
\text{CV}(m_{\rm{stop}}) \leq c_{rC} \times \text{CV}(m_{\rm{stop \_ opt}}).
\end{equation*}
\cite{RobustC} suggested the values $c_{rC} \in \lbrace 1, 1.1, 1.3, 1.5, 2 \rbrace$ for the case of a binary outcome. 
The authors aimed for a less complex predictive rule and for choosing a robust tuning parameter, which is essential in cross-study predictions.

\begin{figure}[t]\centering
\includegraphics[width=0.7\textwidth]{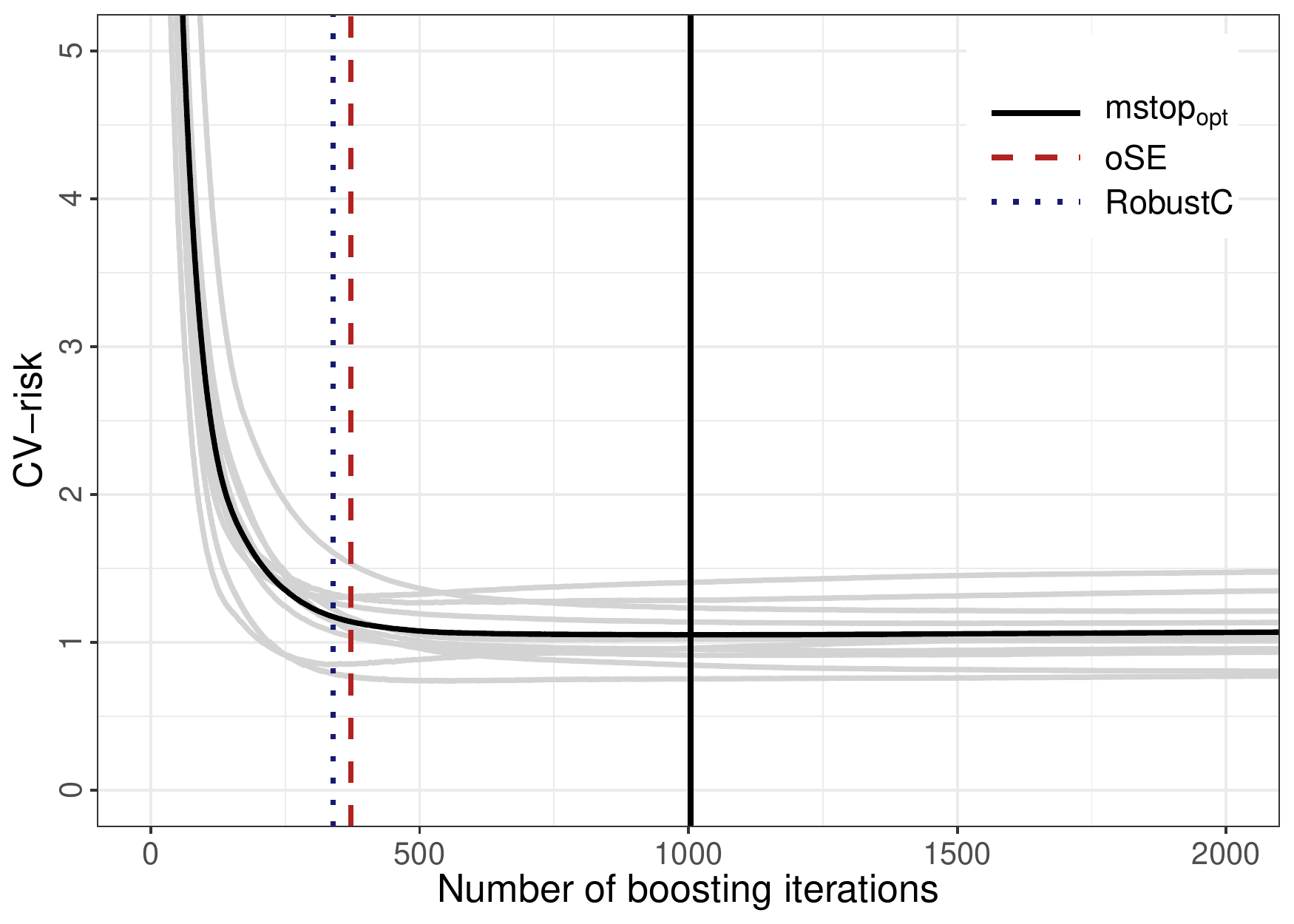}
\caption{Application of the oSE and RobustC on the cross-validated estimation of the empirical risk with 10-fold cross-validation.
The vertical solid black line reflects the optimal stopping iteration via cross-validation ($m_{\rm{stop}} = 1004$), the red dashed one displays the oSE ($m_{\rm{stop}} = 372$) and the blue dotted one RobustC ($m_{\rm{stop}} = 339$) with $c_{rC} = 1.1$. \label{PredictiveRisk}}
\end{figure}

Considering the example from Section~\ref{Intro}, Figure~\ref{PredictiveRisk} shows the CV-risk for 10-fold cross-validation with 2000 boosting iterations.  
The vertical solid black line shows the optimal stopping iteration and corresponds to the minimum average risk over the 10-fold cross-validation samples. The vertical dashed red line is the stopping iteration which yields the oSE, while the blue dotted line corresponds to the optimal iteration according to RobustC. One can observe that the stopping iterations of the earlier stopping strategies are less than half as large as the original $m_{\rm{stop}}$. 

A further alternative approach to obtain sparser models is probing. The idea is to extend the data set by random noise variables, so-called probes, which are randomly shuffled versions of the originally observed variables. 
The algorithm stops when the first probe is selected. For more details on this approach see \cite{probing}.

\subsection{De-selection approach of variables with a small risk reduction}\label{deselection}
Several other approaches have been developed to enhance the sparsity of boosting models (e.g. \citealt{stab_boostI}). Most of them focus on the selection step in the algorithm, or on the tuning of the stopping iteration $m_{\text{stop}}$ (Section~\ref{ES}). Our new procedure is based on \textit{actively} deselecting variables that have been selected by the algorithm, but result in only minor importance regarding the predictions of the model.

We address this issue with an approach that aims at eliminating variables with a small impact and directly enforce the sparsity of the model. The general idea is to first apply a standard boosting algorithm with early stopping via cross-validation or resampling techniques; then, we determine the variables selected by boosting with a minor importance for the model and deselect those components. Afterwards, we boost again incorporating  only the selected variables that survived as candidate variables. In this context, our procedure shows analogies to the twin boosting approach~\citep{twin}.
In our deselection procedure, we consider the risk reduction as a measure for variable importance and deselect those variables that only represent a small percentage of the total risk reduction.

The risk reduction by base-learner $j$ after $m_{\rm{stop}}$ boosting iterations can be defined as the \textit{attributable} risk reduction $R_j$
\begin{equation} \label{Eq:RiskReduction}
R_j = \sum_{m=1}^{m_{\rm{stop}}}I(j = j^{*[m]}) (r^{[m-1]} - r^{[m]}), \quad j = 1,\dots, p,
\end{equation}
where $I$ denotes the indicator function and $ j^{*[m]}$ is the selected base-learner in iteration $m$. Furthermore,  $r^{[m-1]} - r^{[m]}$ represents the risk reduction in iteration $m$, for risks $r^{[m]}$ and $r^{[m-1]}$ at iterations $m$ and $m-1$.
For a given threshold $\tau \in (0,1)$, we deselect base-learner $j$ if
\begin{equation}\label{Eq:CumRisk}
	R_{j} <  \tau \cdot (r^{[0]} - r^{[m_{\rm{stop}}]}),
\end{equation}
where $r^{[0]} - r^{[m_{\rm{stop}}]}$ represents the total risk reduction and $R_{j}$ denotes the attributable risk reduction of base-learner $j$. 

 \begin{figure}
    \centering
    \includegraphics[width = 0.7\textwidth]{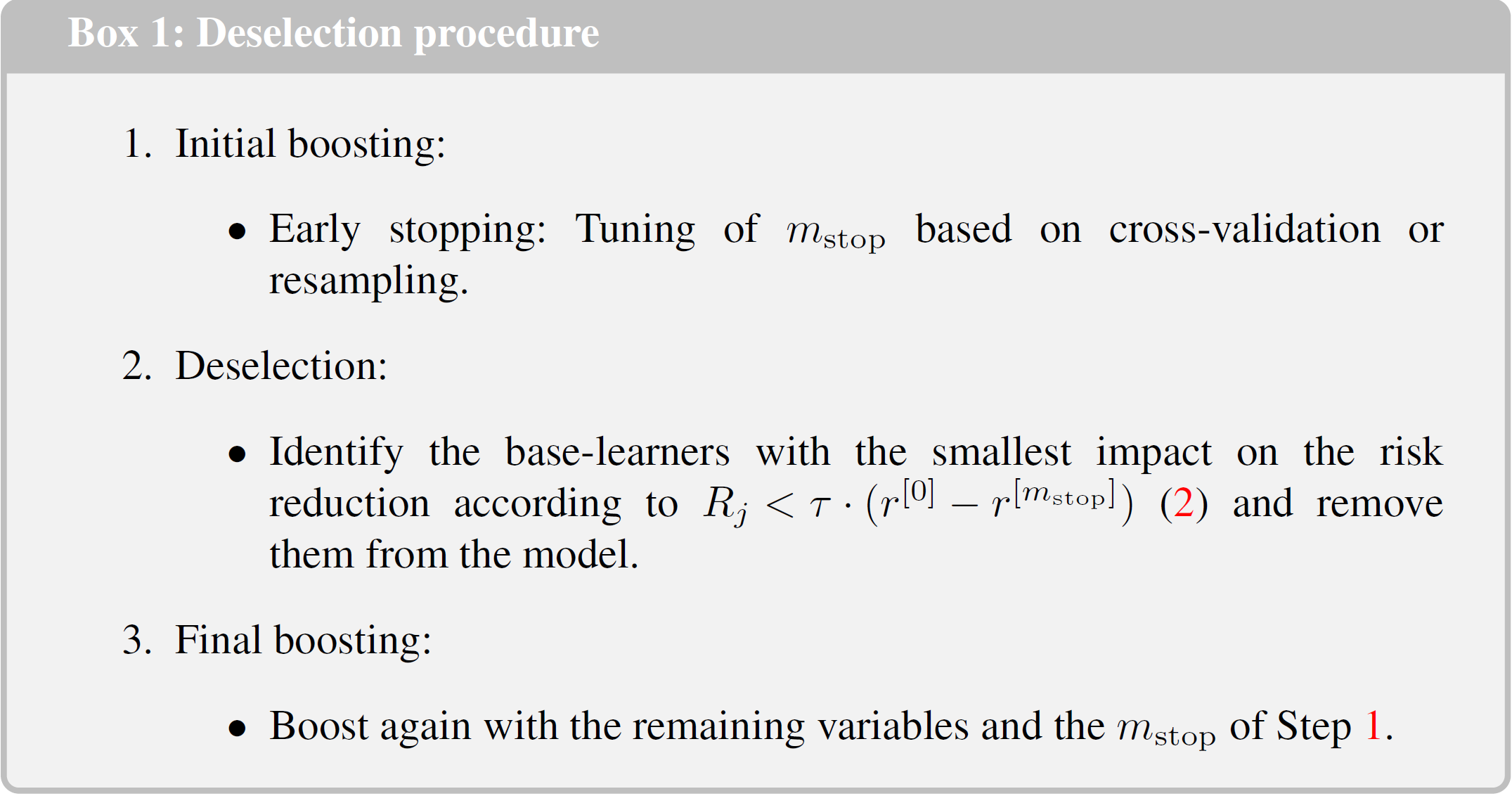}
\end{figure} 

\begin{figure}[t]\centering
\includegraphics[width=\textwidth]{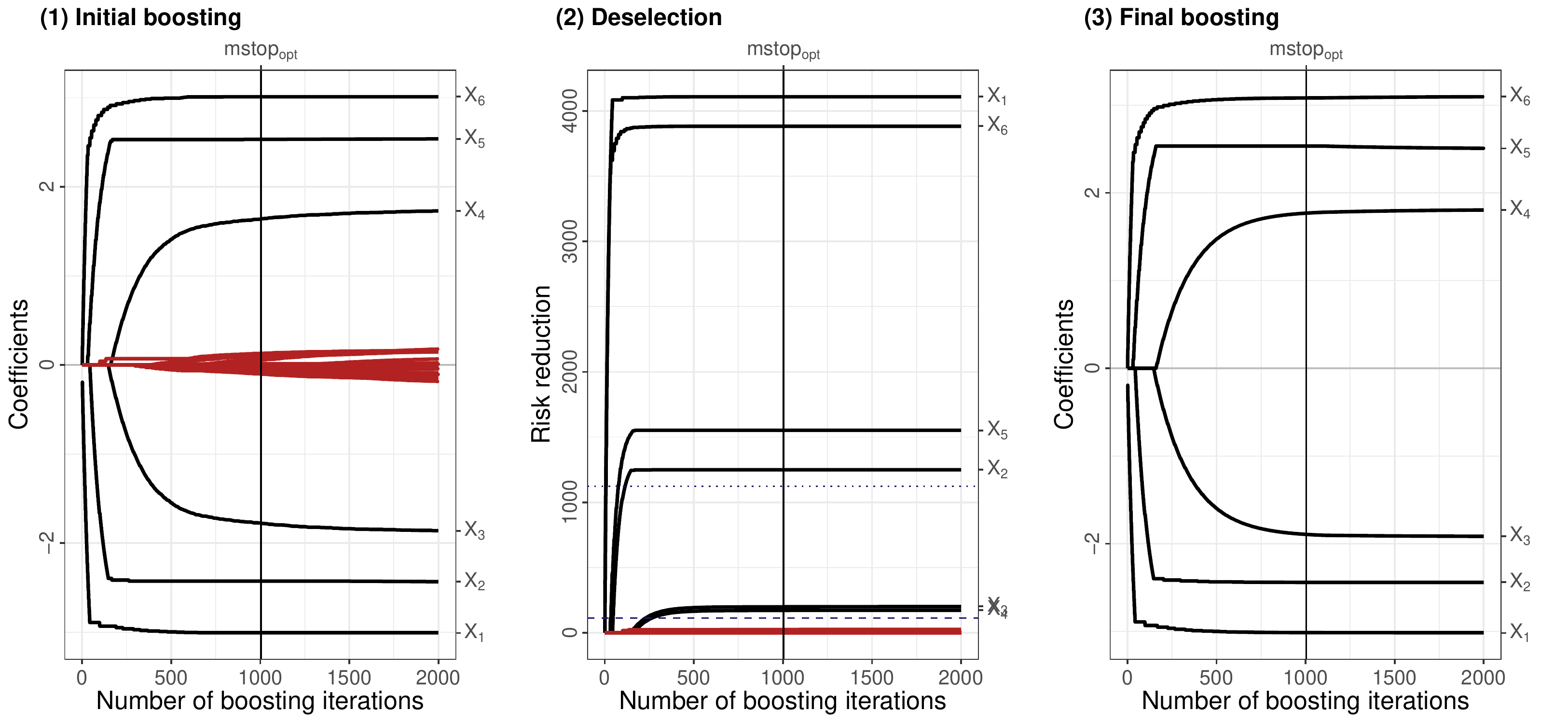}
\caption{Simulation example for the deselection procedure (see Box~1). The left plot shows the coefficient paths of the initial boosting. The central plot displays the attributable risk reduction for the individual variables, together with the 1\% threshold (dashed line) and 10\% threshold (dotted line) of the total risk reduction. The coefficient paths of the right plot correspond to the final boosting. \label{Figure:AttRiskRed}}
\end{figure}

A schematic overview of the proposed procedure can be found in Box~1. Step~1 of the procedure consists of the initial boosting for which the coefficient paths are shown in Figure~\ref{Figure:AttRiskRed} (left), corresponding to the simulation example discussed earlier (see Figures~\ref{Figure:Path_Risk},\ref{PredictiveRisk}). Overall, $23$ variables (of the 50 variables) were selected (shown as horizontal red and black paths) after $m_{\rm{stop}} = 1004$ boosting iterations which were tuned by 10-fold cross-validation (indicated by the vertical black line). For the deselection in Step~2, the attributable risk reduction along the iterations is shown for each individual base-learner in the central plot of Figure~\ref{Figure:AttRiskRed}. To illustrate the effect of the deselection step of the proposed method, consider the thresholds  \(\tau=0.01\) (horizontal dashed line) and \(\tau=0.1\) (horizontal dotted line). Here, it can be observed that our deselection procedure is fundamentally different to earlier stopping approaches discussed in Section~\ref{ES}, as the choice of the threshold for the deselection corresponds to a vertical view on the individual risk reductions after a given number of boosting iterations (see central plot of Figure~\ref{Figure:AttRiskRed}); on the other hand,  earlier stopping simply corresponds to a horizontal shift on the usual regularization paths of boosting (see Figure~\ref{PredictiveRisk}).   

In the following, we consider a threshold value of $\tau = 0.01$ and accordingly deselect those variables which contribute less than 1\% to the total risk reduction. The black paths correspond to the variables included in the model after applying the deselection approach, while the red paths do not cross the 1\% line and the corresponding variables are deselected from the model. 
We can observe that these variables contribute only slightly to the risk reduction and are incorporated with a coefficient size around zero in the initial boosting model (as shown in Figure \ref{Figure:AttRiskRed}, left). In this example, the deselection approach with threshold \(\tau=0.01\) deselects all noise variables from the model, but not the signal variables \(X_1,\dots,X_6\). Variables $X_1, X_2,  X_5$ and $X_6$ have by far the greatest individual contributions to the total risk reduction; however, variables $X_3$ and $X_4$ also exceed the $1\%$ threshold (but not the $10\%$ threshold). After the deselection step, we boost again (Step~3) with only the remaining variables and receive the final model (see Figure \ref{Figure:AttRiskRed}, right) which contains here exclusively the six informative variables.

\subsection{Deselection of base-learners for distributional regression}\label{deselectionGAMLSS}
In classical statistical models, the relationship between a response variable and covariates is most often modeled only based on the expected value. For example, a generalized additive model (GAM)~\citep{GAM} in which the conditional mean $\mu = \mathbb{E}(y|x)$ relates to an additive predictor $\eta$ via a link-function $g$, is given by
\begin{equation*}
g(\mu) = \eta(x)  = \beta_0 + \sum\limits_{j=1}^{p}f_j(x_j)
\end{equation*}
with the intercept $\beta_0$ and the additive effects $f_j$ for the covariates $X_j$ with $j=1,\dots, p$ (including linear, smooth or random effects). Consider for example a Gaussian distribution, which has two parameters: the expected value $\mu$ and the scale parameter $\sigma$. In a classical GAM, we assume that $\sigma$ is fixed and only model the mean parameter $\mu$ in terms of the covariates.

In some cases, this may lead to an overly restrictive point of view, for example in the presence of heteroscedasticity. In addition, skewness and kurtosis may be large so that more complex non-symmetric distributions are required where potentially skewness or higher order moments could be modeled through covariates to obtain a more accurate model.
Following this idea, GAMs have been extended to generalized additive models for location, scale and shape (GAMLSS) by \cite{gamLSS}, where a general parametric density $P(y|\theta_1, \dots \theta_K)$ with distributional parameters $\theta_k$ can be employed.
Here, each distribution parameter $\theta_k$, with $k = 1, \dots, K$, can be modeled by an additive predictor $\eta_{k}$ depending on covariates. Furthermore, for each parameter $\theta_k$, we have parameter-specific link-functions $g_k(.)$ as well as parameter-specific covariates $x_{k1},\dots, x_{kp_k}$.
In general, the linear predictors in a GAMLSS for $K$ distributional parameters can be written as follows:

\begin{equation*}
g_{k}(\theta_k)=\eta_{k}  = \beta_{0k} + \sum\limits_{j=1}^{p_k}f_{jk}(x_{kj}), \quad k = 1, \dots, K,
\end{equation*}
where $\beta_{0 k}$ are the intercepts for the distributional parameters $\theta_k$ and $f_{j k}$ denote the functions of the effect of variable $X_j$ on the parameter $\theta_k$.

GAMLSS can also be fitted via statistical boosting with the package \textbf{gamboostLSS}~\citep{gamboostLSS}. As in the classical setting of boosting GAMs, the main tuning parameter is the stopping iteration $m_{\rm{stop}}$ which controls shrinkage of effect estimates and variable selection. Here, we focus on a non-cyclical boosting approach~\citep{gamboostLSSnoncyclic}, which performs in every iteration only the overall best-performing update among the available candidate variables (base-learners) and distribution parameters. So the term \emph{component-wise} boosting in this context does not only refer to the components of $X$, but also to the components of the parameter space $\theta_1,...,\theta_K$ of the corresponding likelihood. 
To receive the overall best performing base-learner, the empirical risk (the negative log-likelihood) of the best fitted base-learner is determined for each distribution parameter and then compared across the different dimensions.

\begin{figure}[t]\centering
\includegraphics[width=\textwidth]{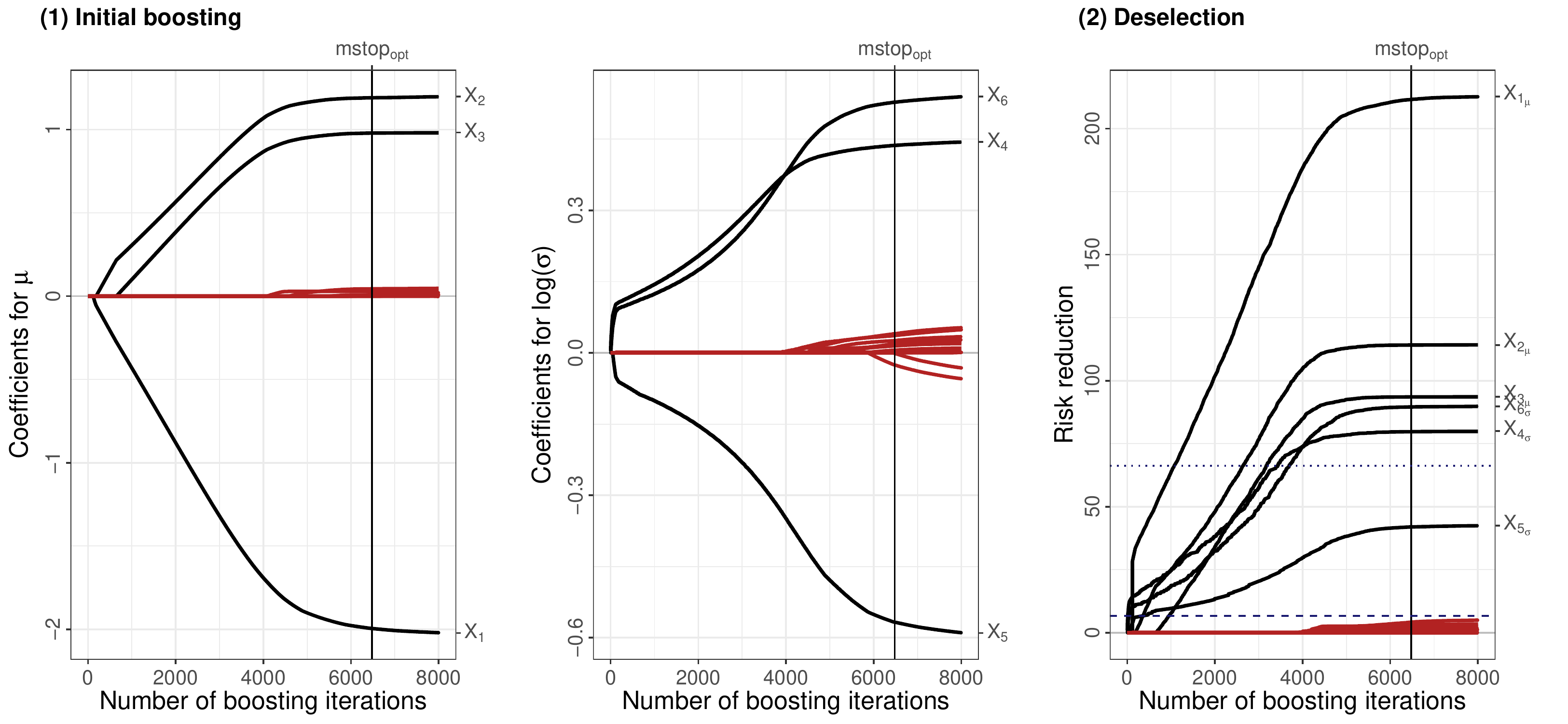}
\caption{Simulation example for gamboostLSS with three informative variables for $\mu$ ($X_{1},X_{2}, X_{3}$) and three for $\sigma$ ($X_{4},X_{5}, X_{6}$). The first two plots display the coefficient paths (for $\mu$ (left) and $\sigma$ (center)) and the third plot shows the attributable risk reductions for the individual variables for both distributional parameters together with the 1\% (dashed) and 10\% line (dotted) of the total risk reduction. The variables corresponding to the black coefficient path are still in the model after deselection with $\tau = 0.01$. \label{Figure:Coef_path_gamboostLSS}}
\end{figure}

The updates are independent for the parameters and each additive predictor may depend on different variables with the guarantee of data-driven variable selection in every submodel: Figure~\ref{Figure:Coef_path_gamboostLSS} displays the estimated coefficient paths in a linear Gaussian location-scale model with distributional parameters $\mu$ (left) and $\sigma$ (center).  
The data set consists of $n = 500$ observations and $p = 20$ variables, where the first three variables $X_{1}, X_{2}, X_{3}$ are informative for the mean parameter $\mu$ with $\beta_{\mu_{\text{inf}}} = (-2, 1.25, 1)^T$, while variables $X_{4}, X_{5}, X_{6}$ are informative for the scale parameter $\sigma$ with $\beta_{\sigma_{\text{inf}}} = (0.5, - 0.5, 0.5)^T$. All other variables are non-informative with $\beta_{\mu} = 0$ and $\beta_{\sigma} = 0$. The explanatory variables were simulated from a multivariate normal distribution with Toeplitz covariance structure and a correlation of $\rho = 0.5$. The optimal number of boosting iterations is $m_{\rm{stop}} = 6479$, optimized via 10-fold cross-validation. Note that in every iteration only a single component is updated for one parameter.

We notice that in the first iterations, components of parameter $\sigma$ were more often updated, which can be observed by the increase of the coefficient sizes for the variables $X_{4}, X_{5}, X_{6}$ in the first iterations.
In total, the boosting model contains 18 of the 50 variables, with $11$ variables selected for $\mu$ and $14$ variables selected for $\sigma$ (where seven variables were selected for both $\mu$ and $\sigma$). Hence, additional variable selection can be advantageous to obtain sparser and thus more interpretable models, which only include the informative variables.

Using equation \eqref{Eq:RiskReduction} and considering $j = 1,\dots, \sum p_k$, the risk reduction in a GAMLSS for component $j$ can be defined similar as before.
For the deselection of variables with a low impact on the risk reduction for distributional regression, we consider the distributional parameters together, where each parameter can depend on different variables. Analogous to equation \eqref{Eq:CumRisk}, we deselect component $j$ if
\begin{equation*}
	R_{j} < \ \tau \cdot (r^{[0]} - r^{[m_{\rm{stop}}]})
\end{equation*}
with fraction $\tau \in (0,1)$ and total risk reduction $r^{[0]} - r^{[m_{\rm{stop}}]}$.
Note that the deselected components may arise from different distributional parameters and that with this definition, GAMs are included as a special case in the general formulation for a GAMLSS with $p_k\equiv p$ and $k=1$. 

For the simulation example, the risk reduction of the variables for $\mu$ and $\sigma$ is shown in the right plot (Figure~\ref{Figure:Coef_path_gamboostLSS}). As in Figure~\ref{Figure:AttRiskRed}, the threshold value is chosen as $\tau = 0.01$ (horizontal dashed line) and $\tau = 0.1$ (horizontal dotted line). The black paths correspond to the variables remaining in the model after applying the deselection procedure (with $\tau = 0.01$) for distributional regression and have by far the highest impact on the risk reduction. The deselection results in a model including only the six informative variables (instead of the 18 initially selected variables). 
For the choice of an appropriate value for the threshold parameter $\tau$, we examined different potential values observing the attributable risk reduction of the base-learner as in Figure~\ref{Figure:AttRiskRed} (second plot) and Figure~\ref{Figure:Coef_path_gamboostLSS} (third plot).

Considering Figure~\ref{Figure:AttRiskRed}, the variables $X_1, X_2, X_5$ and $X_6$ have the largest impact on the risk reduction. All of those variables remain in the model with a deselection threshold of 1\% as well as the other two informative variables $X_3$ and $X_4$. For the 10\% boundary, $X_3$ and $X_4$ would not enter the model because of a smaller risk reduction.
Even for the data example in Figure~\ref{Figure:Coef_path_gamboostLSS}, 10\% is not an appropriate choice, since the variables $X_1, \dots, X_6$ have a noticeable impact on the risk reduction, but $X_5$ would fall out at this limit.
A threshold of 1\% appears to be reasonable in the considered situation. However, in non-sparse situations, when many base-learners contribute only with a small risk reduction to the model, multiple signal variables may be deselected with a threshold of $\tau = 0.01$. This extreme scenario should be rare, and in such non-sparse data situations, enforcing variable selection might not be favorable in general.

An implementation of the enhanced variable selection approach of Section \ref{deselection} and \ref{deselectionGAMLSS} is available at GitHub (\url{https://github.com/AnnikaStr/DeselectBoost}).

\section{Simulation study}\label{Simulation}
To evaluate the performance of our new approach for different data settings, we conduct a simulation study focusing on the variable selection properties as well as the prediction accuracy in comparison with the methods for earlier stopping, described in Section \ref{ES}.

Specifically, the questions to be investigated in the simulation study are as follows:
\begin{enumerate} 
\item Is the direct deselection approach able to identify the truly informative variables (decreasing the number of false positive variables selected by classical boosting)?
\item How does the reduction in selected variables affect the prediction accuracy?
\item How does the new procedure perform in comparison to the earlier stopping strategies, e.g. oSE and RobustC?
\item What is an appropriate value for $\tau$ in the proposed deselection approach? 
\end{enumerate}

\subsection{Settings}
To examine those questions, different settings are considered: First, we start with classical mean regression models (linear, non-linear and logistic regression) and afterwards, we consider the deselection approach in the context of distributional regression models. 

For all simulations, the explanatory variables $X_1,\dots, X_p$ were simulated from a multivariate normal distribution $N(\mathbf{0},\mathbf{\Sigma})$ with a Toeplitz covariance structure $\Sigma_{ij} = \rho^{|i-j|}$ for $1 \leq i, j \leq p$, where $\rho \in (0,1)$ is the correlation between consecutive variables $X_j$ and $X_{j+1}$. For an alternative block-wise covariance structure, see the corresponding results 
in Supplementary Material~A.1.
Overall, we considered two different dimensions of the data problem: i) a low-dimensional setting ($p<n$) with $n = 500$ observations and $p = 20$ variables and ii) a high-dimensional setting ($p>n$) with $n = 500$ observations and $p = 1000$ variables. In total, six of the included variables were informative (for the distributional regression, three for each parameter). Furthermore, a low-correlated scenario with $\rho = 0.2$ and a high-correlated scenario with $\rho = 0.8$ was considered for each setting. 
Additionally, we consider a variation of signal-to-noise ratios (SNRs) and the corresponding effect of different threshold values~$\tau$ with $\text{SNR}\in \lbrace 0.15, 6, 14.64\rbrace$  and $\tau \in \lbrace 0.005,0.0075, 0.01, 0.025, 0.05,0.075, 0.1,0.125\rbrace$.  

For evaluation, we generated test data sets with $1000$ observations from the same distribution as the training data sets. As in the illustrative examples, the number of boosting iterations was tuned via 10-fold cross-validation. The fixed step size is set to $\nu = 0.1$ and was not varied in the simulation, considering that it does not largely affect the risk reduction as long as the step size is chosen reasonably small~\citep{schmid2008boosting}.
We additionally compared the deselection approach (with $\tau=0.01$) with the earlier stopping strategies, oSE and RobustC (additional comparison with probing is given in Supplementary Material A.2). The parameter value for RobustC is chosen as $c_{rC} = 1.05$ for a continuous outcome variable and $c_{rC} = 1.1$ for a binary response, following the recommendation of \cite{RobustC}.

For each setting, 100 simulation runs were conducted and the data sets were generated from the following models:

\begin{enumerate}[left= 0pt,label=\textbf{Scenario~\Alph*},ref={Scenario~\Alph*},wide = 0pt]
 \item (Linear regression)\label{ModelA}\\ 
The true linear model for the continuous outcome variable $Y$ is given by 
\begin{equation*}	
y = -2  x_1  -1.5   x_2 - x_3 + x_4 + 1.5 x_5 + 2   x_6 + \epsilon,
\end{equation*}
with $\epsilon \sim N(0,1)$. The base-learners correspond to simple linear models and the performance was assessed using the mean squared error of prediction (MSEP).
 
\item (Non-linear regression)\label{ModelB}\\ The outcome variable $Y$ was generated from the model
\begin{equation*}	
y =  1.5 \sin(x_1) + x_2 - 0.25 x_3^2 - 0.25x_4 - x_5 -1.5  x_6  + \epsilon,
\end{equation*}
with $\epsilon \sim N(0,1)$. Smooth P-splines were used as base-learners and the MSEP was used for evaluation.

\item (Logistic regression)\label{ModelC} \\ The logistic regression model for covariates with only linear effects on the response was simulated according to
\begin{equation*}
\log \left( \frac{\mathbb{P}(y=1|x)}{\mathbb{P}(y=0|x)} \right) = -5   x_1 + -2.5  x_2 - x_3 + x_4 + 2.5  x_5 + 5 x_6.
\end{equation*}
As evaluation criteria, the Brier score and the Area under the Curve (AUC) were analyzed on test data.

\item (Distributional regression)\label{ModelD}\\ 
For the distributional regression model we consider a Gaussian regression with expected value $\mu$ and scale parameter $\sigma$. Both parameters depend on three different covariates 
\begin{align*}
\mu& = -2  x_1 + 1.25  x_2 +  x_3,\\
\log(\sigma) & = 0.5  x_4 - 0.5 x_5 + 0.5 x_6. \\
\end{align*}
The boosting model was configured with simple linear models as base-learners and the performance was evaluated via the negative log-likelihood.
\end{enumerate}

All simulations were conducted in the statistical computing environment \textbf{R} \citep{R} using the add-on package \textbf{mboost}~\citep{Rmboost} for model-based boosting. The algorithm for fitting GAMLSS models via component-wise gradient boosting is implemented in \textbf{gamboostLSS}~\citep{gamboostLSS}. Twin boosting is implemented in the package \textbf{bst}~\citep{bst}. 
The \textbf{R} code to reproduce the following simulation results can be found online at GitHub (\url{https://github.com/AnnikaStr/DeselectBoost}).

\subsection{Results}
\begin{figure}[t!]\centering
\includegraphics[width=\textwidth]{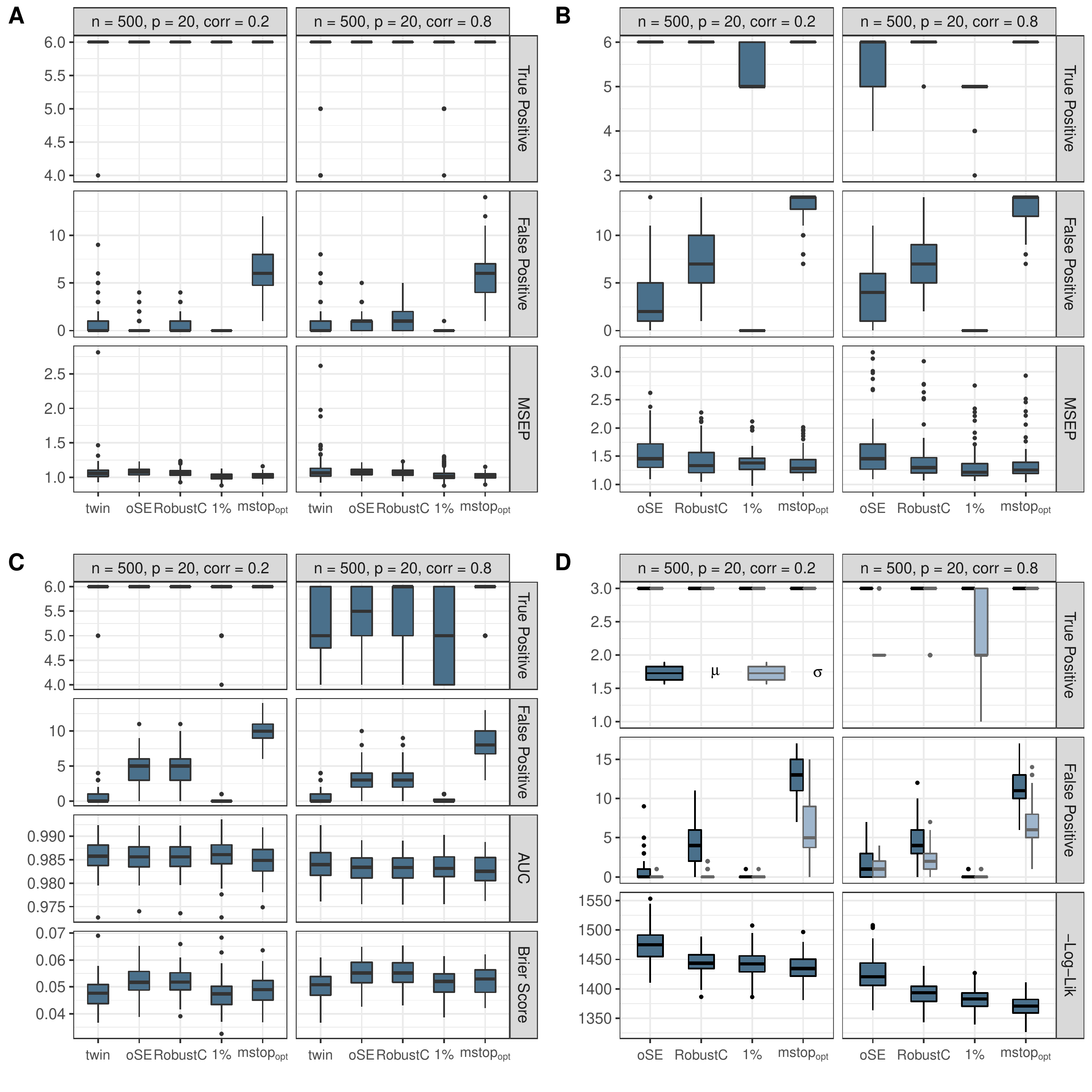}
\caption{Low-dimensional setting: Comparison of the oSE, RobustC, the deselection approach with $\tau = 0.01$ (1\%) and the classical boosted model with $m_{\text{stop\_opt}}$ regarding the true positives, false positives and the prediction performance for A: Linear Regression, B: Non-linear Regression, C: Logistic Regression and D: Distributional Regression. Additional comparisons with twin boosting (twin) for the linear and logistic regression scenarios. \label{Simulation:ResultsLow}}
\end{figure}

Figure~\ref{Simulation:ResultsLow} shows the  results of the low-dimensional simulations regarding the four previously described models for the low- as well as the high-correlated settings, respectively.
For each setting, the true positives, false positives, and the predictive performance for the respective model are shown for the earlier stopping strategies, the deselection procedure with $\tau =0.01$ and the classical boosted model.

In general, the two main strategies (earlier stopping and deselection) resulted in a reduction of false positives. 
In each of the four models, the fewest false positives were obtained with the proposed deselection procedure; more precisely, almost all false positives were deselected. 
For some models, one can observe that the selection of informative variables was slightly influenced by the earlier stopping and deselection approach, particularly for the high-correlated settings. 

In comparison with classical boosting, the deselection procedure yielded comparable or slightly better predictive performances for simulated data based on \ref{ModelA}, \ref{ModelB} and \ref{ModelC}. Furthermore, the earlier stopping strategies usually performed not as well as our approach. Only the AUC in \ref{ModelB} is very similar. For distributional regression (\ref{ModelD}), the classical approach yielded the best results concerning the negative log-likelihood but contained a lot of non-informative variables for both distributional parameters. The deselection approach reduced the false positives almost completely and had only a slightly worse prediction performance.

\begin{figure}[t!]\centering
\includegraphics[width=\textwidth]{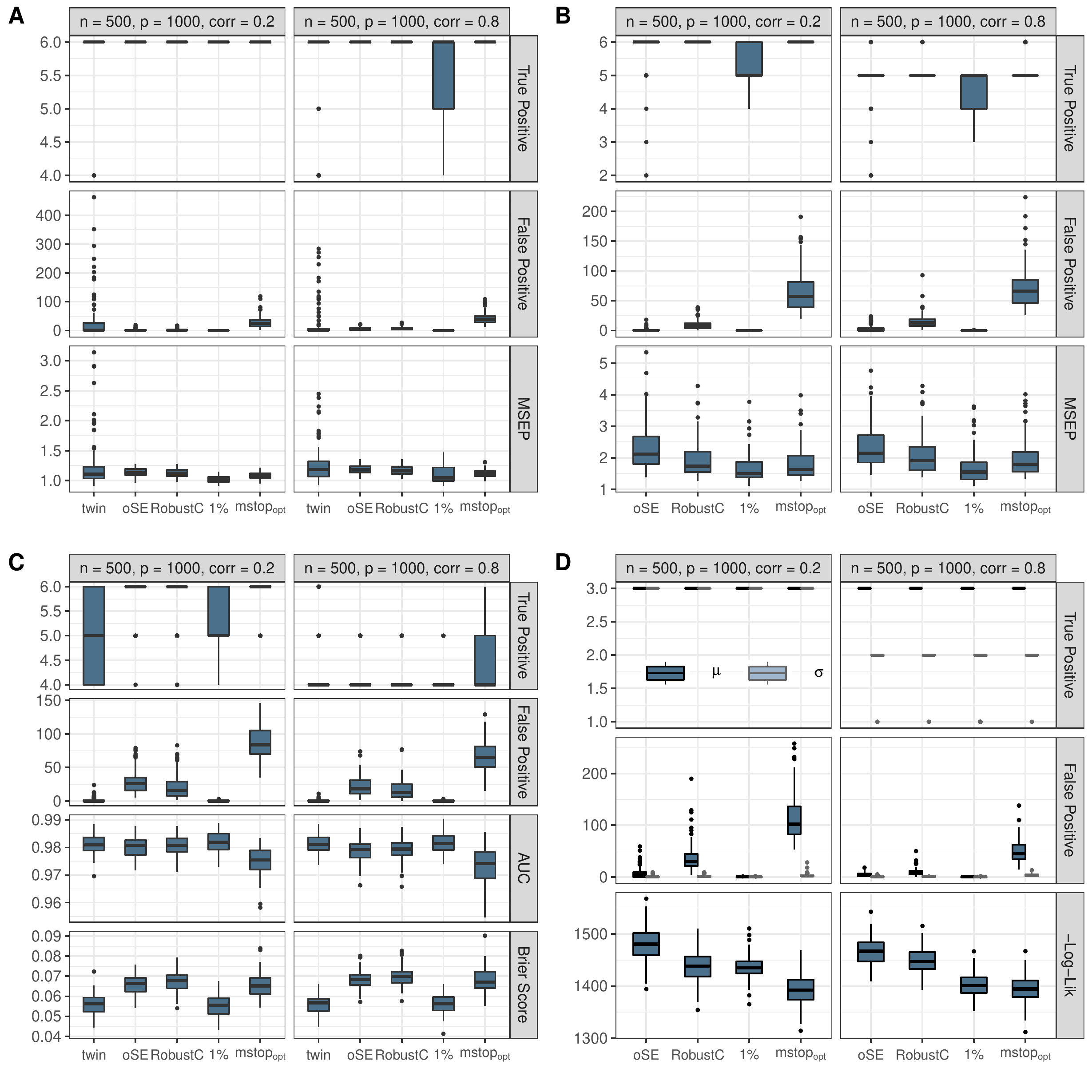} 
\caption{High-dimensional setting: Comparison of the oSE, RobustC, the deselection approach with $\tau = 0.01$ (1\%) and the classical boosted model with $m_{\text{stop\_opt}}$ regarding the true positives, false positives and the prediction performance for A: Linear Regression, B: Non-linear Regression, C: Logistic Regression and D: Distributional Regression. Additional comparisons with twin boosting (twin) for the linear and logistic regression scenarios. \label{Simulation:ResultsHigh}}
\end{figure}

Figure~\ref{Simulation:ResultsHigh} presents the results of the high-dimensional setting. As in Figure~\ref{Simulation:ResultsLow}, the true positives, false positives, and predictive performances are shown. 
For the high-correlated cases of \ref{ModelB}, \ref{ModelC} and \ref{ModelD}, the classical boosting model had already difficulties to select all informative variables. Concerning \ref{ModelC} only four of the six true positives were selected on average. In comparison with the classical approaches, the earlier stopping and deselection approaches resulted in an average lower number of true positives.
For the false positives, we can observe a noticeable reduction with the earlier stopping strategies, but the number of false positives reduced even more with the deselection procedure and the final models contained almost only informative variables. The greatest reduction can be observed for \ref{ModelD} where the classical approach contains 100 false positives on average for parameter $\mu$. After applying the deselection approach, the number of false positives decreased to almost zero with all informative variables still present.
Due to the strong reduction of non-informative variables, in most cases, the deselection procedure showed a better predictive performance in comparison to earlier stopping and the classical boosting. 
Although in most of the simulation runs of \ref{ModelC}, not all informative variables were selected by the proposed deselection approach, it yielded a significantly lower Brier score and a better discriminatory power.  

Furthermore, we compared the new deselection approach as well as earlier stopping strategies to twin boosting in the context of linear and logistic regression models (see  Figure~\ref{Simulation:ResultsLow} and \ref{Simulation:ResultsHigh}). Considering the results for twin boosting of \ref{ModelA}, the number of false positives was reduced (as for oSE, RobustC, and the new deselection procedure), but it shows larger variability, particularly in the high-dimensional setting. 
In one bootstrap sample, the model contained about $450$ false positives (for low correlation). That is much more than we observed with the classical boosting approach, which had the maximum at about $100$ selected non-informative variables. However, it should be noted that these different results for twin boosting are related to a different implementation. The highest decrease in false positives was observed for the deselection approach. 
The prediction accuracy was influenced by the outliers and also showed some higher MSEP values for twin boosting.
We obtained the best model for the deselection approach regarding the number of false positives as well as predictive performance.
The results for logistic regression showed a slight reduction of the selected informative variables for each approach.

On average, twin boosting contained fewer false positives and had a better prediction accuracy than the earlier stopping strategies. Compared with the new deselection procedure, twin boosting tended to include more false positives, but was similar in terms of predictive performance. Here, twin boosting showed favorable properties in comparison to the linear regression model.

\begin{figure}[t!]\centering
\includegraphics[width=\textwidth]{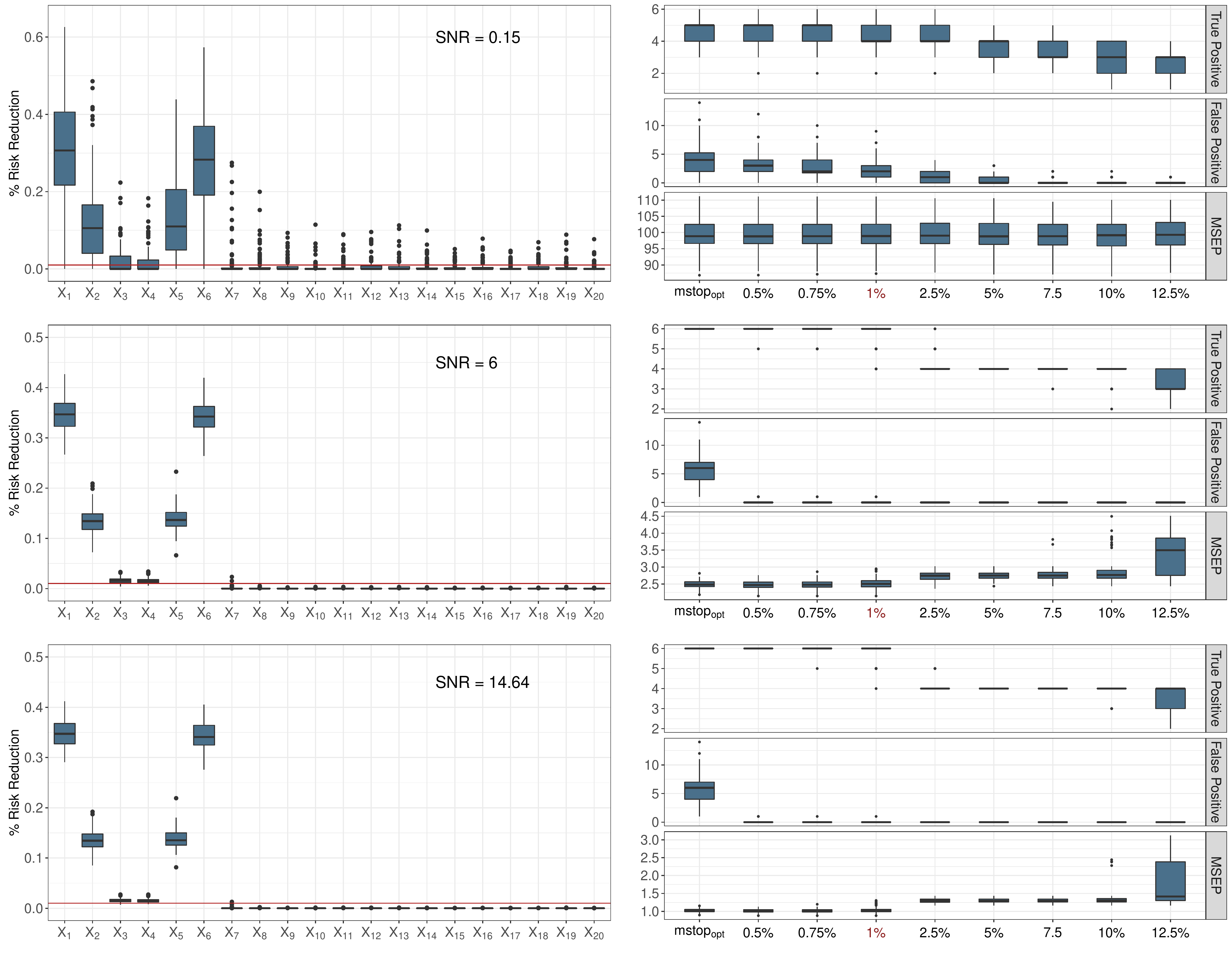}
\caption{Relative risk reduction (in \%) of each base-learner (left) and the variable selection and predictive performance for different $\tau$ values (right) for the low-dimensional setting with correlation $\rho = 0.8$ for increasing signal-to-noise ratios (from top with SNR = 0.15 to bottom with SNR = 14.64) for \ref{ModelA}.   \label{Figure:SNR}}
\end{figure}

Finally, we investigate how the signal-to-noise ratio (SNR) affects the choice of the threshold parameter $\tau$. Figure \ref{Figure:SNR} shows the results for \ref{ModelA} concerning different SNRs with $\text{SNR}\in \lbrace 0.15, 6, 14.64\rbrace$.  
In this case, we consider the low-dimensional setting (for illustrative purposes) with a correlation of 0.8, where SNR = 14.64 corresponds to the simulation setting presented before (results for a correlation of 0.2 and the high-dimensional settings are given in Supplementary Material A.3). 
On the left side, the relative risk reduction (in \%) is depicted for each base-learner for the three different SNRs. The red horizontal line represents a threshold~$\tau$ of 1\%. The right side shows the corresponding results of the true positives and false positives as well as the predictive performance for various $\tau$ values.
Overall, the relative risk reduction for all SNR values was very similar and the highest values always referred to informative variables, of which $X_3$ and $X_4$ showed the lowest risk reduction. The risk reduction for a SNR of 0.15 varied more and the non-informative variables showed a higher contribution to the risk reduction.

Considering the variable selection and predictive performance for various $\tau$ thresholds, the true positives and false positives for a SNR of 6 and 14.64 were very similar over the different $\tau$ values. For a SNR of 0.15, the classical boosting model had larger difficulties to identify all informative variables. 
Hence, it is also more challenging to deselect the non-informative variables without a further reduction in the true positives. Therefore, smaller $\tau$ values are more appropriate, causing less deselection and more noise variables, but the signal variables still remain in the model. 
For the other SNRs, only variable~$X_7$ contributes to the risk reduction for small threshold values (0.5\%, 0.75\%, and 1\%). Here, the relative risk reduction showed that a small $\tau$ value is sufficient to remove almost all false positives (red horizontal line for 1\%). 
Furthermore, a higher $\tau$ value can lead not only to a reduction in informative variables included in the model but also to a worse MSEP. Due to the noise, the predictive performance for the SNR of 0.15 was very poor and showed no discernible differences between the threshold values. However, $\tau$ values above 2.5\% lead to a decrease in performance for larger SNRs. Furthermore, the previous simulation results have shown that a low value for $\tau$ (in this case 1\%) reduces the number of false positives and additionally leads to a comparable predictive performance to classical boosting.

Overall, the number of false positives in the resulting models could be significantly reduced by earlier stopping or deselection as well as twin boosting compared to classical boosting.  However, in most cases, the reduction of false positives for oSE and RobustC resulted in worse prediction performance. A comparison with probing for \ref{ModelA}, \ref{ModelB} and \ref{ModelC} showed similar behavior (given in Supplementary Material A.2). Probing also led to a reduction in the number of false positives, but resulted in worse prediction performance, particularly for \ref{ModelB}. Furthermore, the earlier stopping strategies removed a few informative variables from the model in some settings.\\
The new procedure also deselected some informative variables from time to time, but removed the non-informative variables almost completely and resulted in favorable prediction performance.  In some settings, the new approach even resulted in better predictive performance than the classical boosting model.
Additional simulation results for the high-dimensional setting with a block structure for the covariance matrix are provided in Supplementary Material A.1 and showed very similar results compared with the Toeplitz covariance structure.\\
From the consideration of different SNRs, we conclude that the relative risk reduction attributed to a base-learner does not depend much on the overall signal-to-noise ratio but on the distribution of the signal among the base-learners. To ensure that not too  many  informative variables are de-selected and to achieve a favorable predictive performance, our results suggest that the threshold value should be chosen rather small (e.g., 1\%). For larger SNR values (6 and 14.64), almost all noise variables were eliminated even for small $\tau$. Higher threshold values resulted in worse performance and a significant reduction of the informative variables.  The choice of 1\% for the threshold~$\tau$ resulted in a reasonable trade-off between sparsity and prediction performance in all considered settings. The \textit{best} threshold, however, will always depend on the actual goal of the analysis and the general data situation.

Additionally, our approach also performed well compared to twin boosting, particularly for the linear regression model. An advantage of our method is the possibility to enhance variable selection for non-linear and  distributional regression, which to the best of our knowledge is currently not available for twin boosting.

\section{Quality of life of chronic kidney disease patients}\label{GCKG}

The following analysis aims to identify the most important predictors for the quality of life of stage~III chronic kidney disease patients based on an ongoing German cohort study (German Chronic Kidney Disease Study, GCKD). A similar analysis has already been published (c.f.,~\cite{betaboost}) and led to the selection of rather large models which partly motivated the current new methodological developments. 

The analysis is based on beta regression~\citep{betareg}, which is a very flexible approach to model bounded outcome variables  like proportions. It is also a well-known tool in the analysis of health-related quality of life scores~\citep{hunger2011analysis,hunger2012longitudinal} which typically range from 0 (lowest possible value) to 100 (highest possible value).
The density function of a beta distribution with expected value $\mu$ and  precision parameter $\phi$ is given by
\begin{equation*}
	f(y; \mu, \phi) = \frac{\Gamma(\phi)}{\Gamma(\mu\phi)\Gamma((1-\mu)\phi)} y^{\mu\phi-1}(1-y)^{(1-\mu)\phi-1},  \quad 0<y<1,
\end{equation*}
where $\Gamma(.)$ denotes the gamma function. In context of distributional beta regression, which refers to a generalized additive model for location, scale and shape (GAMLSS), we model $\mu$ and additionally $\phi$ in terms of several explanatory variables.

\begin{table}[t]
\small\sf\centering
\caption{Results for GCKD data in terms of the mean (sd) number of selected variables for the parameters $\mu$ and $\phi$ as well as the negative log-likelihood representing the prediction performance on the 1000 bootstrap replicates.\label{Table:Illustration}}
\begin{tabular}{lcccccc}
\toprule
Model & $\mu$ & $\phi$ &  $-$log-likelihood \\ 
\midrule
classical boosted model & 26.43 (7.10) & 14.55 (6.00) & -1457.08 (40.24)\\
deselected ($\tau = 0.01$) & 12.58 (1.39) & 7.87 (1.61)  & -1441.73 (39.41)\\
oSE &  8.06 (2.68) & 2.92 (1.63)  & -1295.34 (37.88)\\
RobustC & 7.63 (2.31) & 2.70 (1.34)  & -1290.14 (32.88)\\
\bottomrule
\end{tabular}
\end{table}

The GCKD study~\citep{eckardt2012german} is an ongoing cohort study for patients with stage~III chronic kidney disease. We analyzed part of the cross-sectional baseline-data with $n=3522$ observations and 54 explanatory variables. We aimed to select the most informative variables for the quality of life of chronic kidney disease patients~\citep{betaboost} using the \textbf{R} add-on package \textbf{betaboost}.
The effects of the predictors on the quality of life are represented by base-learners. For continuous covariates we incorporated spline effects as base-learners. For factor variables (for example, education and exercise) we used linear base-learners providing joint updates of the effects for the different categories in the boosting iterations. Therefore, our approach yields potential deselection (sparsity) on the full factor level and not on the level of different categories of a factor. Alternatively, multi-categorical factors may also be re-coded as several binary dummy variables, so that categories could be selected (and deselected) independently.

\begin{figure}[b!]\centering
\includegraphics[width=\textwidth]{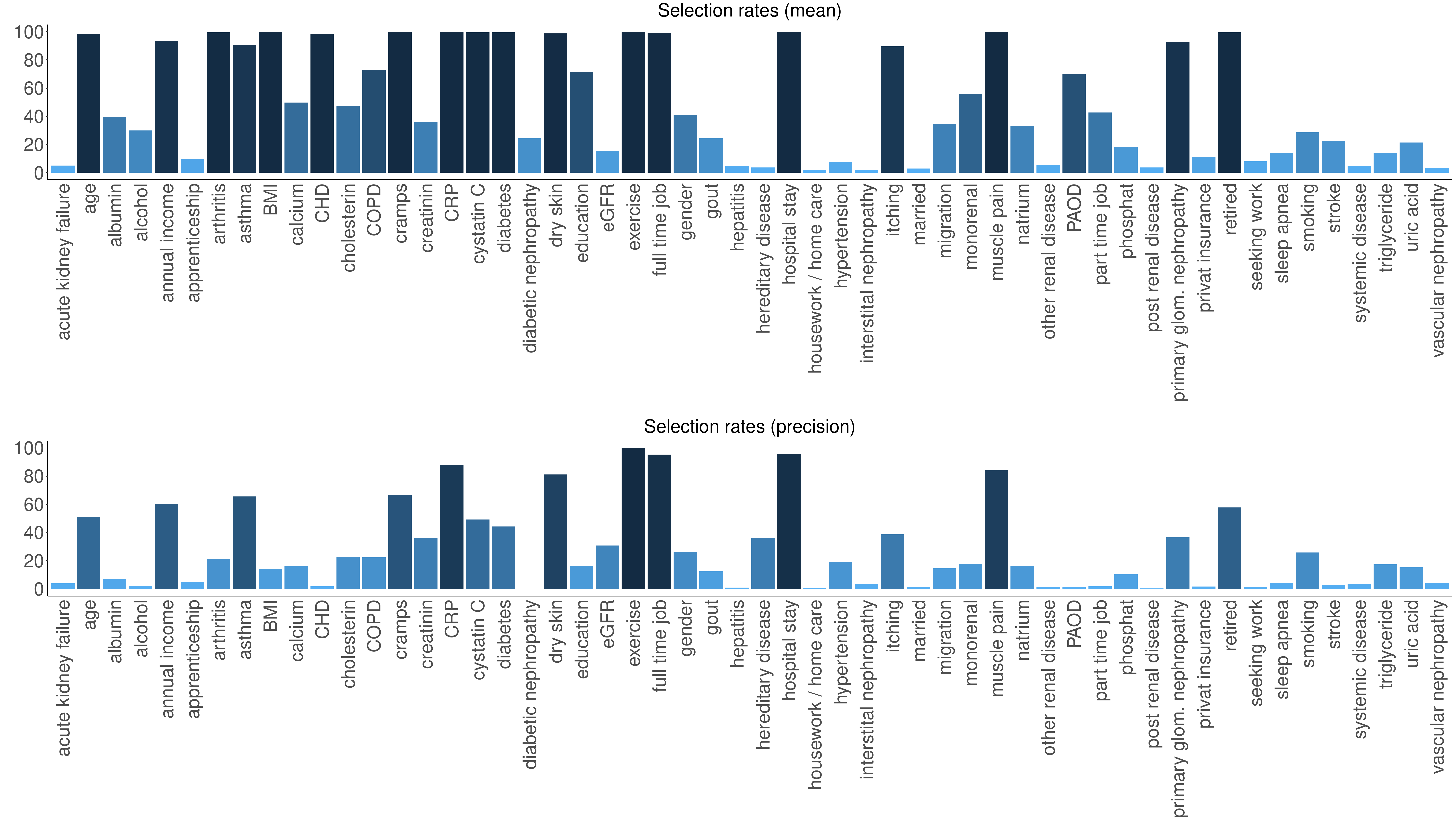}
\caption{The selection rates of the explanatory variables for $\mu$ and $\phi$ of the classical boosting algorithm in 1000 bootstrap samples.\label{SelRateClassic}}
\end{figure}

We drew 1000 bootstrap replicates and fitted a beta regression model without and with the new deselection procedure using $\tau=1\%$ for each bootstrap sample (results for different $\tau$ values are given in Supplementary Material A.4). To evaluate the predictive performance of the resulting models, the negative log-likelihood was computed on the ``out-of-bag'' bootstrap samples.
The optimal number of boosting iterations were selected via 10-fold cross-validation. For comparison, we additionally considered the oSE and RobustC methods.

\begin{figure}[t!]\centering
\includegraphics[width=\textwidth]{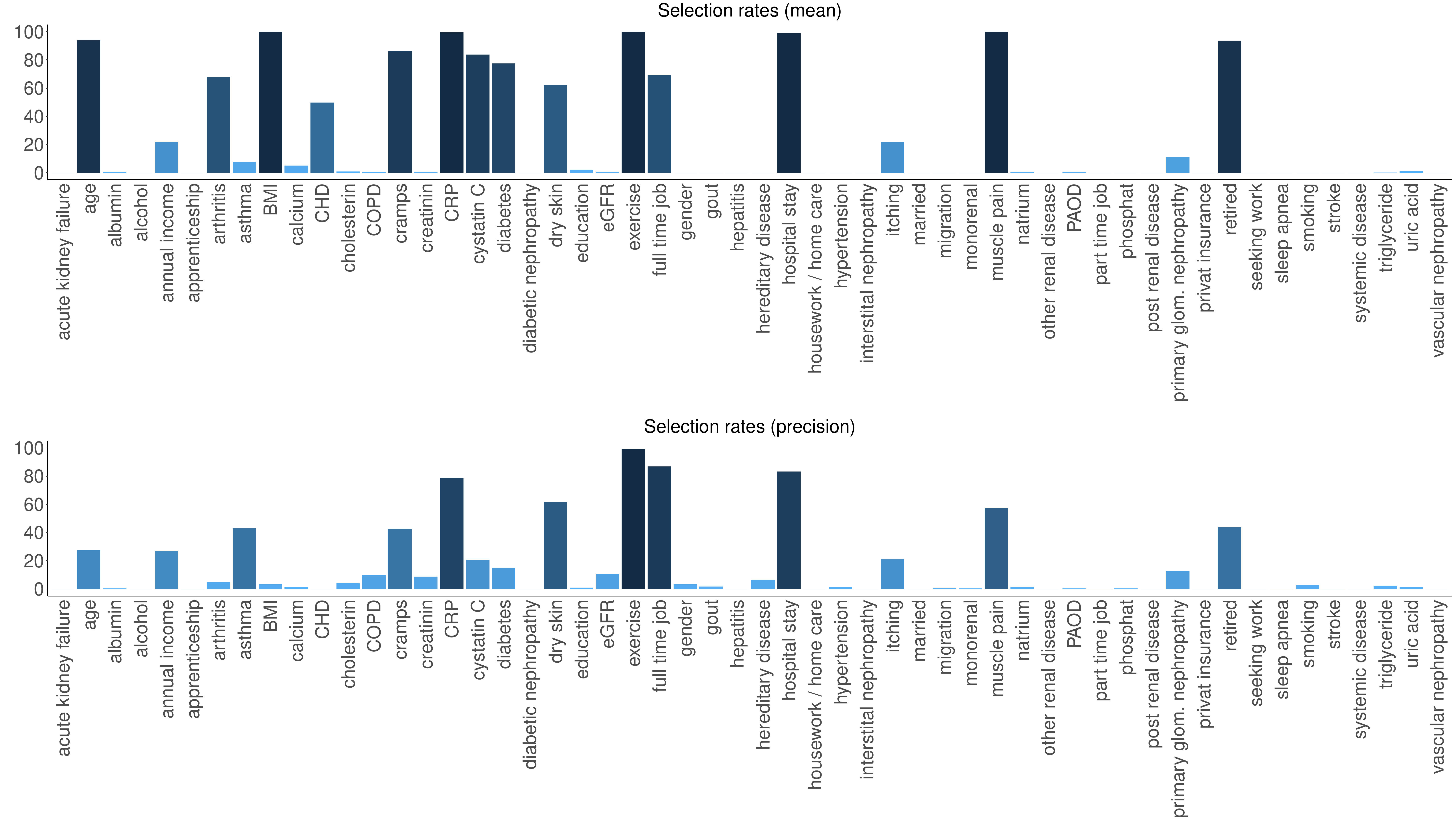}
\caption{The selection rates of the explanatory variables for $\mu$ and $\phi$ after applying the new deselection approach with $\tau = 0.01$ in 1000 bootstrap samples.\label{SelRateTau}}
\end{figure}

\subsection{Results}
Table \ref{Table:Illustration} displays the mean numbers (with standard deviations) of selected variables for $\mu$ and $\phi$ as well as the average negative log-likelihood for the different models on the 1000 bootstrap replicates. One can observe that more variables are included for the expected value than for the precision parameter. The earlier stopping strategies contain fewer variables than the proposed deselection approach for boosting.

In addition to Table~\ref{Table:Illustration}, we consider the selection rates for each variable (for $\mu$ and $\phi$) on the 1000 bootstrap replicates. Figure~\ref{SelRateClassic} displays the selection rates of the classical boosting approach. As described in \cite{betaboost}, the highest selection rates for parameter $\mu$ were obtained for age, body mass index (BMI), exercise, and variables related to pain such as arthritis, cramps and muscle pain.
Furthermore, variables that are indicators of kidney failure and inflammation also had higher rates, e.g., cystatin C. 
For the precision parameter $\phi$, 15 variables were included on average, with the highest rates for the variables exercise, employment in a full-time job and hospital stay.

The selection rates after additionally applying the deselection approach in Figure~\ref{SelRateTau} show that the new procedure achieved a significant reduction in the number of included variables; some variables that were rarely selected by classical boosting were never included with the new approach (e.g. alcohol, gender). On the other hand, the variables with the highest selection rates from the classical model were still present at the highest selection rates.

\begin{figure}[t]\centering
\includegraphics[width=0.7\textwidth]{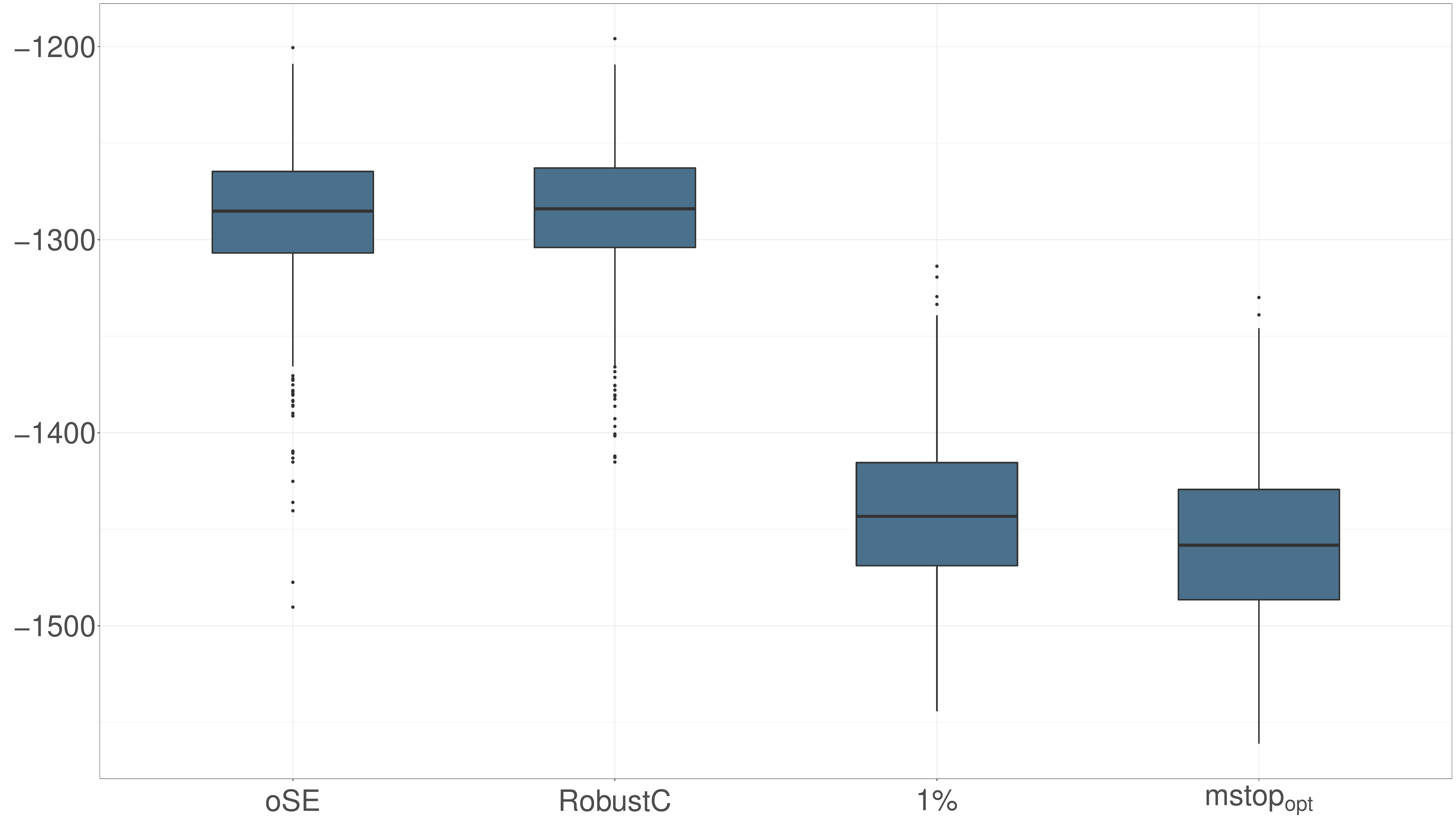}
\caption{Negative log-likelihood of the oSE, RobustC, the deselection procedure with a threshold value of $1\%$ and the classical boosted model on ``out-of-bag'' bootstrap samples.\label{Illustration:Likelihood}}
\end{figure}

To evaluate the predictive performance of the resulting models, we considered the negative log-likelihood on test data as a scoring rule. The results in Figure~\ref{Illustration:Likelihood} suggest that the new deselection procedure outperforms the earlier stopping strategies oSE and RobustC. The smallest negative log-likelihood was obtained with the classical boosting model with an average value of 1457.08 (see Table \ref{Table:Illustration}), whereby we achieved a comparable performance for deselecting with $\tau = 0. 01$.

Overall, the deselection approach based on 1\% of the total risk reduction was able to enhance the sparsity of boosting models by selecting less predictors for the  health-related QoL in chronic kidney disease patients (on average 12.6 for the expected value and 7.9 for the precision parameter from 54 candidate variables) in comparison to the classical boosting approach. However, many predictors contribute to the overall risk reduction to a small extent (see Supplementary Material, Figure A7). That indicates that the "true" underlying model is not as sparse as in the simulations. Earlier stopping strategies can further increase the sparsity, however leading to much poorer predictive performance on ``out-of-bag'' data (Figure~\ref{Illustration:Likelihood}). 
In comparison, the deselection approach leads to a slightly worse predictive performance in comparison to classical boosting, but yields much smaller and more interpretable models. \\
In practice, we have to deal with the trade-off between sparsity and predictive performance, which is regulated by the threshold parameter $\tau$. Therefore, higher variable deselection (i.e., larger values of $\tau$) leads to smaller models but, at least in this considered application, also to poorer predictive performance.  

\section{Discussion \& Conclusions}\label{Discussion}

The presented approach to deselect base-learners for enhanced variable selection in statistical boosting is a new technique to obtain sparser models with simpler interpretation via the removal of irrelevant predictors with negligible impact.
As the deselection is based on the risk reduction, this approach is suitable for any type of base-learners, for example, linear models, splines, and spatial effects. Furthermore, the deselection approach can also be combined with a wide range of regression settings, including multi-dimensional optimization problems like distributional regression. 

Compared to the similar twin-boosting~\citep{twin}, our approach actively deselects base-learners via a threshold value. This is somehow an analogy to stability selection~\citep{stabilityI, stabilityII} but our method focuses on providing a sparse prediction model instead of a set of stable predictors~\citep{stab_boostI, stab_boostII}. Furthermore, it does not include additional resampling steps. Other approaches for enhanced variable selection in the context of boosting focused on strategies for earlier stopping~\citep{probing, RobustC} which typically also increases the amount of shrinkage on effect estimates which might not be necessarily favourable. Our approach hence focuses on a vertical view on the regularization paths (in contrast to the horizontal view with earlier stopping) which has already been discussed in the literature on the lasso~\citep{zhou2009thresholding, weinstein2020power}.

The new approach is particularly suitable for high-dimensional data (with more potential predictors than observations) as one can obtain a simplified model with the most relevant variables yielding in many cases almost the same prediction accuracy as the classical boosting approach without deselection. Consequently, the interpretability of resulting prediction models improves, which makes their application in practice more likely~\citep{wyatt1995commentary}.

The results of the simulation study suggest that our procedure can yield much sparser models by deselecting wrongly selected variables; in many cases deselection was associated with an almost complete elimination of false positives. In practice, one could assume that this might often lead also to a decreased prediction accuracy: the standard boosting approach already selects the optimal prediction model by optimizing the stopping iteration. However, at least in some of the simulation settings, the deselection of false positives led even to a slightly improved prediction performance. 

To select the most informative predictors for the health-related quality of life of chronic kidney disease patients (GCKD study), the deselection procedure resulted in a drastically reduced set of variables compared to a recent analysis~\citep{betaboost}, which partly motivated the new methodological development. However, we did also observe a slight worsening of the model performance (w.r.t. the likelihood on test data).

The deselection procedure is controlled via a threshold value $\tau$: it represents the minimum amount of total risk reduction which should be attributed to a corresponding base-learner in order to avoid deselection. In the simulation study, a threshold of $\tau = 0.01$ (i.e., 1$\%$ of total risk reduction) was considered to be appropriate overall.
However, the general trade-off between a more complex model with the highest possible prediction accuracy and a sparser, more interpretable model (higher \emph{descriptive accuracy}~\citep{Murdoch22071} with potentially reduced prediction accuracy) should be guided by the researcher, depending on the research question and the context of the problem. 
As an alternative, the threshold parameter could also be chosen via resampling techniques or cross-validation which might further increase the performance but leads also to higher computational burden, particularly for high-dimensional data.

A limitation of our procedure is the assumption of sparsity. In cases where this is not fulfilled, it might deselect too many variables: If, for example, many predictors affect the model with minor impact (e.g., 200 variables with equal importance), our approach with $\tau = 0.01$  may deselect all variables. This is due to the dependency of our approach on the distribution of risk reduction across the base-learners. In theory, it would be beneficial to select $\tau$ based on the minimal signal strength, e.g., the minimal risk reduction attributed to an informative predictor. As the truly informative variables, however, are unknown -- this  choice remains challenging in practical applications. An alternative technique, particular for non-sparse settings, could be to consider the \textit{cumulative} risk reduction. Instead of considering only the risk reduction attributed to the corresponding base-learner, the cumulative risk considers the risk reduction of all base-learners that are to be deselected from the model. Thus, this procedure accounts for the complete tail of the base-learners with low importance. This variant  would typically yield larger models when used with the same threshold. We investigated also the deselection via the cumulative risk reduction (results for the simulation and the application are given in Supplementary Material B). This alternative version is also implemented and available together with the corresponding code to reproduce the simulations and can be applied by specifying \texttt{method = "cumulative"}.

The favorable performance of our new approach motivates research in this direction in the future, in particular for distributional regression~\citep{gamlssbook}, where sparse and interpretable models are of particular importance.  For instance, deselection could be also extended to the level of distribution parameters in order to deselect a complete model-dimension (e.g., decreasing a GAMLSS to a GAM) when the contribution to the overall risk reduction is limited. 
Another line of potential research could focus on the combination of earlier stopping with deselection to avoid the disadvantage of exaggerated shrinkage~\citep{van2020regression}.

Altogether, we conclude that  in our simulation and application the new deselection approach was able to outperform existing methods for earlier stopping, concerning the number of selected variables and the predictive performance. However, it should be noted that these approaches pursue various different goals like variable selection, prediction performance and/or interpretability. Fitting one model, that is able to achieve the best solution for all potential goals, simply might often not be feasible (cf.,~\citealt{HothornDiscStabSel}). 

\subsection*{Acknowledgment}
We thank Benjamin Hofner for fruitful discussions on the underlying methodology of the new deselection procedure. 

\subsection*{Funding}
The work on this article was supported by the Deutsche Forschungsgemeinschaft (DFG, grant number 428239776, KL3037/2-1, MA7304/1-1). The  GCKD  study  was  funded  by  grants  from  the  German Ministry  of  Education  and  Research  (BMBF)  (\url{http://www.gesundheitsforschung-bmbf.de/de/2101.php}; grant number 01ER0804) and the  KfH  Foundation  for  Preventive  Medicine  (\url{http://www.kfh-stiftung-praeventivmedizin.de/content/stiftung}).

\bibliographystyle{apalike}
\bibliography{Deselection}

\begin{thebibliography}{}

\bibitem[Breiman et~al., 1984]{breiman}
Breiman, L., Friedman, J., Stone, C.~J., and Olshen, R.~A. (1984).
\newblock {\em Classification and Regression Trees}.
\newblock CRC Press, Boca Raton.

\bibitem[B{\"u}hlmann et~al., 2014]{buhlmann2014discussion}
B{\"u}hlmann, P., Gertheiss, J., Hieke, S., Kneib, T., Ma, S., Schumacher, M.,
  Tutz, G., Wang, C.-Y., Wang, Z., Ziegler, A., et~al. (2014).
\newblock Discussion of “the evolution of boosting algorithms” and
  “extending statistical boosting”.
\newblock {\em Methods of Information in Medicine}, 53(6):436--445.

\bibitem[Bühlmann and Hothorn, 2007]{buhlmann}
Bühlmann, P. and Hothorn, T. (2007).
\newblock Boosting algorithms: Regularization, prediction and model fitting.
\newblock {\em Statistical Science}, 22(4):477--505.

\bibitem[Bühlmann and Hothorn, 2010]{twin}
Bühlmann, P. and Hothorn, T. (2010).
\newblock Twin boosting: improved feature selection and prediction.
\newblock {\em Statistics and Computing}, 20:119--138.

\bibitem[Chen et~al., 2020]{jasa_prs}
Chen, T.-H., Chatterjee, N., Landi, M.~T., and Shi, J. (2020).
\newblock A penalized regression framework for building polygenic risk models
  based on summary statistics from genome-wide association studies and
  incorporating external information.
\newblock {\em Journal of the American Statistical Association}, 0(0):1--11.

\bibitem[Choi et~al., 2020]{choi2020tutorial}
Choi, S.~W., Mak, T. S.-H., and O’Reilly, P.~F. (2020).
\newblock Tutorial: a guide to performing polygenic risk score analyses.
\newblock {\em Nature Protocols}, 15(9):2759--2772.

\bibitem[Eckardt et~al., 2012]{eckardt2012german}
Eckardt, K.-U., B{\"a}rthlein, B., Baid-Agrawal, S., Beck, A., Busch, M.,
  Eitner, F., Ekici, A.~B., Floege, J., Gefeller, O., Haller, H., et~al.
  (2012).
\newblock The german chronic kidney disease ({GCKD}) study: design and methods.
\newblock {\em Nephrology Dialysis Transplantation}, 27(4):1454--1460.

\bibitem[Ellenbach et~al., 2020]{RobustC}
Ellenbach, N., Boulesteix, A., Bischl, B., Unger, K., and Hornung, R. (2020).
\newblock Improved outcome prediction across data sources through robust
  parameter tuning.

\bibitem[Fan and Lv, 2010]{HighDimData}
Fan, J. and Lv, J. (2010).
\newblock A selective overview of variable selection in high dimensional
  feature space.
\newblock {\em Statistica Sinica}, 20(1):101--148.

\bibitem[Ferrari and Cribari-Neto, 2004]{betareg}
Ferrari, S. and Cribari-Neto, F. (2004).
\newblock Beta regression or modeling rates and proportions.
\newblock {\em Journal of Applied Statistics}, 31(7):799--815.

\bibitem[Freund, 1990]{Boosting_ML}
Freund, Y. (1990).
\newblock Boosting a weak learning algorithm by majority.
\newblock In {\em Proceedings of the Third Annual Workshop on Computational
  Learning Theory}, COLT '90, page 202–216, San Francisco, CA, USA. Morgan
  Kaufmann Publishers Inc.

\bibitem[Freund and Schapire, 1996]{FreundShapire}
Freund, Y. and Schapire, R.~E. (1996).
\newblock Experiments with a new boosting algorithm.
\newblock In {\em Proceedings of the Thirteenth International Conference on
  International Conference on Machine Learning}, ICML'96, page 148–156.
  Morgan Kaufmann Publishers Inc.

\bibitem[Friedman, 2001]{friedman2001}
Friedman, J. (2001).
\newblock Greedy function approximation: A gradient boosting machine.
\newblock {\em The Annals of Statistics}, 29(5):1189--1232.

\bibitem[Friedman et~al., 2000]{friedman2000}
Friedman, J., Hastie, T., and Tibshirani, R. (2000).
\newblock Additive logistic regression: a statistical view of boosting (with
  discussion and a rejoinder by the authors).
\newblock {\em The Annals of Statistics}, 28(2):337--407.

\bibitem[Friedman et~al., 2010]{glmnet}
Friedman, J., Hastie, T., and Tibshirani, R. (2010).
\newblock Regularization paths for generalized linear models via coordinate
  descent.
\newblock {\em Journal of Statistical Software}, 33(1):1--22.

\bibitem[Hastie et~al., 2009]{hastie2009elements}
Hastie, T., Tibshirani, R., and Friedman, J. (2009).
\newblock {\em The Elements of Statistical Learning: Data Mining, Inference,
  and Prediction}.
\newblock Springer series in statistics. Springer.

\bibitem[Hofner et~al., 2015]{stab_boostI}
Hofner, B., Boccuto, L., and G{\"o}ker, M. (2015).
\newblock Controlling false discoveries in high-dimensional situations:
  boosting with stability selection.
\newblock {\em BMC Bioinformatics}, 16(1):144.

\bibitem[Hofner et~al., 2014]{RTutorial}
Hofner, B., Mayr, A., Robinzonov, N., and Schmid, M. (2014).
\newblock Model-based boosting in {R}: A hands-on tutorial using the {R}
  package mboost.
\newblock {\em Computational Statistics}, 29:3--35.

\bibitem[Hothorn, 2010]{HothornDiscStabSel}
Hothorn, T. (2010).
\newblock Invited discussion on {``Meinshausen and B\"uhlmann: Stability
  Selection''}.
\newblock {\em Journal of the Royal Statistical Society: Series B (Statistical
  Methodology)}, 72(4):463--464.

\bibitem[Hothorn et~al., 2020]{Rmboost}
Hothorn, T., Buehlmann, P., Kneib, T., Schmid, M., and Hofner, B. (2020).
\newblock {\em {mboost}: Model-Based Boosting}.
\newblock {R} package version 2.9-2.

\bibitem[Hunger et~al., 2011]{hunger2011analysis}
Hunger, M., Baumert, J., and Holle, R. (2011).
\newblock Analysis of {SF-6D} index data: is beta regression appropriate?
\newblock {\em Value in Health}, 14(5):759--767.

\bibitem[Hunger et~al., 2012]{hunger2012longitudinal}
Hunger, M., D{\"o}ring, A., and Holle, R. (2012).
\newblock Longitudinal beta regression models for analyzing health-related
  quality of life scores over time.
\newblock {\em BMC Medical Research Methodology}, 12(1):144.

\bibitem[Mayr et~al., 2014a]{mayr2014evolution}
Mayr, A., Binder, H., Gefeller, O., and Schmid, M. (2014a).
\newblock The evolution of boosting algorithms.
\newblock {\em Methods of Information in Medicine}, 53(6):419--427.

\bibitem[Mayr et~al., 2014b]{mayr2014extending}
Mayr, A., Binder, H., Gefeller, O., and Schmid, M. (2014b).
\newblock Extending statistical boosting.
\newblock {\em Methods of Information in Medicine}, 53(06):428--435.

\bibitem[Mayr et~al., 2012a]{gamboostLSS}
Mayr, A., Fenske, N., Hofner, B., Kneib, T., and Schmid, M. (2012a).
\newblock Generalized additive models for location, scale and shape for high
  dimensional data—a flexible approach based on boosting.
\newblock {\em Journal of the Royal Statistical Society: Series C (Applied
  Statistics)}, 61:403--427.

\bibitem[Mayr et~al., 2012b]{knowstop}
Mayr, A., Hofner, B., and Schmid, M. (2012b).
\newblock The importance of knowing when to stop a sequential stopping rule for
  component-wise gradient boosting.
\newblock {\em Methods of Information in Medicine}, 51:178--86.

\bibitem[Mayr et~al., 2016]{stab_boostII}
Mayr, A., Hofner, B., and Schmid, M. (2016).
\newblock Boosting the discriminatory power of sparse survival models via
  optimization of the concordance index and stability selection.
\newblock {\em BMC Bioinformatics}, 17(1):288.

\bibitem[Mayr et~al., 2017]{mayr2017update}
Mayr, A., Hofner, B., Waldmann, E., Hepp, T., Gefeller, O., and Schmid, M.
  (2017).
\newblock An update on statistical boosting in biomedicine.
\newblock {\em Computational and Mathematical Methods in Medicine}, 2017:1--12.

\bibitem[Mayr et~al., 2018]{betaboost}
Mayr, A., Weinhold, L., Hofner, B., Titze, S., Gefeller, O., and Schmid, M.
  (2018).
\newblock {The betaboost package - a software tool for modelling bounded
  outcome variables in potentially high-dimensional epidemiological data}.
\newblock {\em International Journal of Epidemiology}, 47(5):1383--1388.

\bibitem[Meinshausen and B{\"u}hlmann, 2010]{stabilityI}
Meinshausen, N. and B{\"u}hlmann, P. (2010).
\newblock Stability selection.
\newblock {\em Journal of the Royal Statistical Society: Series B (Statistical
  Methodology)}, 72(4):417--473.

\bibitem[Murdoch et~al., 2019]{Murdoch22071}
Murdoch, W., Singh, C., Kumbier, K., Abbasi-Asl, R., and B., Y. (2019).
\newblock Definitions, methods, and applications in interpretable machine
  learning.
\newblock {\em Proceedings of the National Academy of Sciences},
  116(44):22071--22080.

\bibitem[{R Core Team}, 2019]{R}
{R Core Team} (2019).
\newblock {\em R: A Language and Environment for Statistical Computing}.
\newblock R Foundation for Statistical Computing, Vienna, Austria.

\bibitem[Rigby and Stasinopoulos, 2005]{gamLSS}
Rigby, R.~A. and Stasinopoulos, D.~M. (2005).
\newblock Generalized additive models for location, scale and shape.
\newblock {\em Journal of the Royal Statistical Society: Series C (Applied
  Statistics)}, 54(3):507--554.

\bibitem[Sauerbrei et~al., 2020]{sauerbrei2020state}
Sauerbrei, W., Perperoglou, A., Schmid, M., Abrahamowicz, M., Becher, H.,
  Binder, H., Dunkler, D., Harrell, F.~E., Royston, P., Heinze, G., et~al.
  (2020).
\newblock State of the art in selection of variables and functional forms in
  multivariable analysis—outstanding issues.
\newblock {\em Diagnostic and Prognostic Research}, 4:1--18.

\bibitem[Schmid and Hothorn, 2008]{schmid2008boosting}
Schmid, M. and Hothorn, T. (2008).
\newblock Boosting additive models using component-wise p-splines.
\newblock {\em Computational Statistics \& Data Analysis.}, 53(2):298--311.

\bibitem[Shah and Samworth, 2013]{stabilityII}
Shah, R.~D. and Samworth, R.~J. (2013).
\newblock Variable selection with error control: another look at stability
  selection.
\newblock {\em Journal of the Royal Statistical Society: Series B (Statistical
  Methodology)}, 75(1):55--80.

\bibitem[Stasinopoulos et~al., 2017]{gamlssbook}
Stasinopoulos, M.~D., Rigby, R.~A., Heller, G.~Z., Voudouris, V., and
  De~Bastiani, F. (2017).
\newblock {\em Flexible regression and smoothing: using GAMLSS in R}.
\newblock CRC Press.

\bibitem[Steyerberg and Vergouwe, 2014]{steyerberg2014towards}
Steyerberg, E.~W. and Vergouwe, Y. (2014).
\newblock Towards better clinical prediction models: seven steps for
  development and an abcd for validation.
\newblock {\em European Heart Journal}, 35(29):1925--1931.

\bibitem[Thomas et~al., 2017a]{probing}
Thomas, J., Hepp, T., Mayr, A., and Bischl, B. (2017a).
\newblock Probing for sparse and fast variable selection with model-based
  boosting.
\newblock {\em Computational and Mathematical Methods in Medicine}.

\bibitem[Thomas et~al., 2017b]{gamboostLSSnoncyclic}
Thomas, J., Mayr, A., Bischl, B., Schmid, M., Smith, A., and Hofner, B.
  (2017b).
\newblock Gradient boosting for distributional regression: faster tuning and
  improved variable selection via noncyclical updates.
\newblock {\em Statistics and Computing}, 28:1--15.

\bibitem[Van~Calster et~al., 2020]{van2020regression}
Van~Calster, B., van Smeden, M., De~Cock, B., and Steyerberg, E.~W. (2020).
\newblock Regression shrinkage methods for clinical prediction models do not
  guarantee improved performance: Simulation study.
\newblock {\em Statistical Methods in Medical Research}, 29(11):3166--3178.

\bibitem[Wang, 2020]{bst}
Wang, Z. (2020).
\newblock {\em bst: Gradient Boosting}.
\newblock {R} package version 0.3-23.

\bibitem[Weinstein et~al., 2020]{weinstein2020power}
Weinstein, A., Su, W.~J., Bogdan, M., Barber, R.~F., and Candès, E.~J. (2020).
\newblock A power analysis for knockoffs with the lasso coefficient-difference
  statistic.

\bibitem[Wood, 2017]{GAM}
Wood, S. (2017).
\newblock {\em Generalized Additive Models: An Introduction with R}.
\newblock Chapman and Hall/CRC, 2 edition.

\bibitem[Wyatt and Altman, 1995]{wyatt1995commentary}
Wyatt, J.~C. and Altman, D.~G. (1995).
\newblock Prognostic models: clinically useful or quickly forgotten?
\newblock {\em British Medical Journal}, 311(7019):1539--1541.

\bibitem[Zhou, 2009]{zhou2009thresholding}
Zhou, S. (2009).
\newblock Thresholding procedures for high dimensional variable selection and
  statistical estimation.
\newblock In Bengio, Y., Schuurmans, D., Lafferty, J.~D., Williams, C.~K., and
  Culotta, A., editors, {\em Advances in Neural Information Processing Systems
  22}, pages 2304 -- 2312, Red Hook, NY. Curran.
\newblock 23rd Annual Conference on Neural Information Processing Systems (NIPS
  2009); Conference Location: Vancouver, Canada; Conference Date: December
  7-10, 2009; Poster presentation.

\bibitem[Zou, 2006]{adaLasso}
Zou, H. (2006).
\newblock The adaptive lasso and its oracle properties.
\newblock {\em Journal of the American Statistical Association},
  101:1418--1429.

\end{thebibliography}

\end{document}